\documentclass[a4paper,fleqn,usenatbib]{mnras}
\usepackage{newtxtext,newtxmath}
\usepackage{breqn}
\usepackage[T1]{fontenc}
\usepackage{ae,aecompl}
\usepackage{tablefootnote}

\usepackage{graphicx}	
\usepackage{amsmath}	

\title[AstroSat view of GRS 1915$+$105]
{{\it AstroSat} view of GRS 1915$+$105 during the Soft State: Detection of HFQPOs and 
estimation of Mass and Spin}

\author[Sreehari et al.]{Sreehari H.$^{1,2}$\thanks{E-mail: hjsreehari@gmail.com, sreehari@iiap.res.in}, 
Anuj Nandi$^{1}$\thanks{E-mail: anuj@ursc.gov.in}, Santabrata Das$^{3}$\thanks{E-mail: sbdas@iitg.ac.in}, V. K. Agrawal$^{1}$, Samir Mandal$^{4}$\thanks{E-mail: samir@iist.ac.in},  
\newauthor M. C. Ramadevi$^{1}$, Tilak Katoch$^{5}$ \\
1. Space Astronomy Group, ISITE Campus, U. R. Rao Satellite Centre, Outer
Ring Road, Marathahalli, Bangalore, 560037, India.\\
2. Indian Institute of Astrophysics, Bangalore, 560034, India. \\
3. Indian Institute of Technology Guwahati, Guwahati, 781039, India.\\
4. Indian Institute of Space Science and Technology, Thiruvananthapuram, 695547, India.\\5. DAA, Tata Institute of Fundamental Research, Colaba, Mumbai, 400005, India.}

\date{Accepted XXX. Received YYY; in original form ZZZ}

\pubyear{2020}

\begin{document}
\label{firstpage}
\pagerange{\pageref{firstpage}--\pageref{lastpage}}
\maketitle

\begin{abstract}

We report the results of {\it AstroSat} observations of GRS 1915$+$105 obtained using  100 ks guaranteed-time (GT) during the soft state. The Color-Color Diagram (CCD) indicates a variability class of $\delta$ with the detection of High Frequency QPO (HFQPO) in the power density spectra (PDS). The HFQPO is seen to vary in the frequency range of $67.96 - 70.62$ Hz with percentage rms $\sim 0.83 - 1.90$ \% and significance varying from $1.63 - 7.75$. The energy dependent power spectra show that the HFQPO features are  dominant only in $6 - 25$~keV energy band. The broadband energy spectra ($0.7 - 50$ keV) of {\it SXT} (Soft X-ray Telescope) and {\it LAXPC} (Large Area X-ray Proportional Counter) modelled with \texttt{nthComp} and \texttt{powerlaw} imply that the source has an extended corona in addition to a compact `Comptonizing corona' that produces high energy emission and exhibits HFQPOs. The broadband spectral modelling indicates that the source spectra are well described by thermal Comptonization with electron temperature (kT$_{\rm e}$) of $2.07 - 2.43$~keV and photon-index ($\Gamma_{\rm nth}$) between $1.73-2.45$ with an additional \texttt{powerlaw} component of photon-index ($\Gamma_{\rm PL}$) between $2.94 - 3.28$. The norm of \texttt{nthComp} component is high ($\sim 8$) during the presence of strong HFQPO and low ($\sim 3$) during the absence of HFQPO. Further, we model the energy spectra with the \texttt{kerrbb} model to estimate the accretion rate, mass and spin of the source. Our findings indicate that the source accretes at super-Eddington rate of $1.17-1.31~ \dot{M}_{\rm Edd}$. Moreover, we find the mass and spin of the source as $12.44 - 13.09~M_{\odot}$ and $0.990-0.997$ with $90\%$ confidence suggesting that GRS 1915$+$105 is a maximally rotating stellar mass X-ray binary black hole source.

\end{abstract}


\section{Introduction}

The astrophysical compact objects like white dwarf (WD), neutron star (NS) and black hole (BH) are often seen to exist along with its companion star. These binary sources are usually X-ray bright and referred as X-ray binaries \citep[XRBs,][]{Seward2010}. The high energy X-ray emission from black hole XRBs (BH-XRBs) is attributed to the accretion processes that are involved in the mass transfer from the companion star to the black hole. Detailed study of the X-ray emission from BH-XRBs is essential to understand the accretion dynamics as well as the effect of strong gravity in the vicinity of the black holes.
 
XRBs are known to exhibit excess power in certain frequencies called Quasi-periodic Oscillations (QPOs) which are evident from the power density spectrum (PDS) of these sources \citep[and references therein]{VanderKlis1985,Belloni2005,Remillard-McClintock06, Nandi-etal2012,Belloni-Motta16,Sreehari-etal2019}. The QPOs in BH-XRBs lie in a wide range of frequencies $\sim 0.1 - 450$ Hz and they have been classified into two general categories: (a) low-frequency QPOs (LFQPOs) in the range $<40$ Hz and (b) high-frequency QPOs (HFQPOs) in the range $\sim 40 - 450$ Hz \citep{Remillard-McClintock06}. The origin of HFQPOs is of great interest as these oscillations are transient as well as subtle. Moreover, HFQPOs possibly are the manifestations of various relativistic effects in the orbits close to the black hole and it can be used as an important tool to probe general relativity in the strong gravity limit \citep[and references therein]{Stella-Vietri98, Rebusco-08, Merloni-etal99, Vincent-etal13, Stefanov14}. However, the conclusive consensus on the origin of HFQPOs is not reached yet.

Some BH-XRBs like GRS 1915$+$105 and IGR J17091$-$3624 also exhibit visually identifiable variabilities in their light curves. These variabilities are also associated with changes in the corresponding energy spectra and are categorized into several variability classes using Color-Color Diagrams (CCDs), light curve profiles \citep{Belloni-etal00} and nature of energy spectra \citep{Iyer2015,Rad2018}. 

Energy spectral modelling of various BH-XRBs indicates that the emission
from these sources are in general thermal as well as non-thermal in nature. 
The multi-temperature disc black body emission represents the Keplerian accretion disc
\citep{ShaSu1973}, whereas the high energy emission from the source 
is attributed due to the inverse-Comptonisation \citep{Sunyaev1980, Tanaka1995, CT1995} of seed 
photons by the hot corona located at the inner part of the disc surrounding the black hole.
Several other models along with the additional features like elemental abundance and 
relativistic-reflection \citep[and references therein]{Garcia2014} have been 
introduced to explain overall emission features. 
However, these models are phenomenological in nature as they do not directly provide the physical properties (namely, mass and spin) of the accreting BH source. Keeping this in mind, in this paper, we attempt to model the broadband energy spectra of GRS 1915$+$105 using the \texttt{kerrbb} \citep{Li2005} model. This model deals with rotating black hole 
and is used to estimate the mass and spin of the sources.

Since detection, the source GRS 1915$+$105 has remained active in X-rays \citep{Castro1992}. However, the source underwent into the non-active phase in the recent past although some flickering activities in X-rays were occasionally seen. It is noteworthy that GRS 1915$+$105  
exhibits fourteen different structured X-ray variability classes \citep{Belloni-etal00, Klein_wolt_etal2002,Hann2005}. In a classical effort, \cite{Greiner-etal01} estimated the mass of the source 
to be $M_{\rm BH} = 14 \pm 4~M_{\odot}$ considering the orbital period of $\sim 33.5$ days. 
Recently, \cite{Reid2014} estimated the black hole mass as $12.4_{-1.8}^{+2.0}$~M$_{\odot}$ and the distance as $8.6_{-1.6}^{+2.0}$~kpc using parallax method. \cite{Zd2014} calculated the inclination of GRS 1915$+$105 to be $\sim 65^{\circ}$ based on kinematic constraints from the mass ejections.
On the other hand, several attempts have been made to estimate the spin
of the source. The estimation of spin of GRS $1915+105$ is reported to be $a_{\rm k}$ = 0.98$^{+0.01}_{-0.01}$ \citep{Blum-etal09, Miller-etal13}.

GRS 1915$+$105 exhibited LFQPOs during the {\it RXTE} era \citep[and references therein]{Ratti2012}. Besides this, \cite{Yadav-etal2016} detected LFQPOs from GRS 1915$+$105 using {\it AstroSat/LAXPC} data. 
GRS 1915$+$105 has also exhibited 67 Hz and 40 Hz HFQPOs during the {\it RXTE} era \citep{Morgan1997, Strohmayer2001}.
\cite{Belloni2006} reported the detection of a 170 Hz HFQPO in the C-State of GRS 1915$+$105.	
Moreover, \cite{Belloni-Altamirano13} detected HFQPOs with frequencies varying from $63-71$ Hz using {\it RXTE} data. \cite{Mendez2013} showed a detailed comparison of the lag spectra for 35 Hz and 67 Hz HFQPOs of GRS 1915$+$105.
Recently, \cite{Belloni2019} reported HFQPOs of frequencies between $67.4-72.3$ Hz in the $\omega$ and $\mu$ variability classes of GRS 1915$+$105 from the {\it AstroSat} observations of July 2016. 
They also studied the phase lags and their relation with hardness ratio, although they ignored the dead-time effect while carrying out power spectral analysis and detection of QPOs. Moreover, they did not carry out the broadband spectral modelling to understand the emission processes. Meanwhile, several models were proposed to explain the origin of the HFQPOs observed in GRS 1915$+$105. \cite{Morgan1997} first proposed that the HFQPOs are associated with the Keplerian frequency at the inner-most stable circular orbit (ISCO). \cite{Cui-etal98} suggested that the HFQPOs possibly arise due to the relativistic Lense-Thirring precession around a highly spinning black hole. \cite{Nowak-etal97} proposed a model based on disko-seismic g-modes of oscillation for explaining the HFQPOs of this source. However, these models bear the limitations of revealing the inherent properties of the source under consideration including the relevant radiative emission processes.

Indeed, the broadband X-ray spectra of GRS 1915$+$105 was studied
\citep{Grove1998,Zd2001} with combined observations from {\it CGRO/OSSE} 
and {\it RXTE}. The spectra from few keV to 10 MeV showed a single {\it powerlaw} with no high energy cut-off up to 500 keV, unlike other BH-XRBs. {\it RXTE} spectra of the source was modelled with just a thermal Comptonization component \citep{Vilhu2001}.
{\it BeppoSAX} spectra of B and C states were found to have a definite high-energy cut-off, varying from 45 to 100 keV \citep{Feroci1999}.

Keeping all these in mind, in this paper, we use Guaranteed Time (GT) {\it AstroSat} data of June 2016 to identify the variability class and characterize the detected HFQPOs of GRS 1915$+$105. We carry out the energy dependent study of
power spectral features to understand the HFQPO properties.
Combining the data from both imaging ({\it SXT}) and the proportional counter units ({\it LAXPC}) on board {\it AstroSat}, we generate and fit the broadband energy spectra with phenomenological as well as physical models in order to get an insight of the disc emission processes. Moreover, we attempt to constrain the mass and spin of the source using broadband spectral modelling and finally compute the mass accretion rate as well.

We organize the paper as follows. In \S \ref{s:obs}, we present the reduction and extraction procedures of {\it AstroSat} data. In \S \ref{s:timing}, we discuss the methods of timing analysis and present the results. In \S \ref{s:spec}, we carry out the spectral analysis and present the results of broadband spectral modelling with {\it AstroSat} data. In \S \ref{s:kerr}, we attempt to constrain the mass and spin of the black hole including the accretion rate. Finally, in \S \ref{s:disc_grs}, we discuss the results of our spectro-temporal studies of the BH source GRS 1915$+$105 and conclude.

\section{Observation and Data Reduction}

\label{s:obs}

The source GRS 1915$+$105 was observed during the period of June 11, 2016 (MJD 57550)
to June 15, 2016 (MJD 57554) using Guaranteed Time (GT) of {\it AstroSat} \citep{Agarwal-etal17} for $100$ ks (Observation ID: G05\_189T01\_9000000492).
In this paper, we make use of the data from the imaging instrument {\it SXT} and the proportional counter {\it LAXPC} on board {\it AstroSat}. {\it SXT} \citep{Singh2017} operates in the $0.3 - 8$~keV energy range.
{\it LAXPC} \citep{Antia-etal17} consists of three proportional counters ({\it LAXPC10, LAXPC20} and {\it LAXPC30}) 
operating in the energy range of $3 - 80$ keV with a high temporal
resolution of $10~\mu$s \citep{Yadav-etal2016}. 

{\it SXT} data reduction is carried out using the software provided by SXT-POC\footnote{\url{http://www.tifr.res.in/~astrosat\_sxt/sxtpipeline.html}} at TIFR. Once the cleaned event files are generated using the pipeline, we use \textsc{XSELECT} tool to extract images, light curves and spectra. However, as the {\it SXT} time resolution is poor (0.28~s), we do not use it for temporal analysis. 
For spectral modelling, we use the redistribution matrix files (rmfs), ancillary response files (arfs) and the background spectral files provided by the {\it SXT} instrument team at TIFR\footnote{\url{http://www.tifr.res.in/~astrosat\_sxt/dataanalysis.html}}.

We use the software available at the {\it AstroSat} science support cell\footnote{\url{http://astrosat-ssc.iucaa.in/?q=data_and_analysis}} for the extraction of light curves of source as well as background from the {\it LAXPC} data.
The response files and software for data extraction are provided by {\it LAXPC} instrument team of TIFR\footnote{\url{http://www.tifr.res.in/~astrosat\_laxpc/LaxpcSoft.html}}.
Further, we follow the procedures as described in \cite{Agrawal-etal2018} and \cite{Sreehari-etal2019} 
to extract energy spectra from both {\it SXT} and {\it LAXPC} observations. 
Once the light curves and energy spectra are generated, we carry out temporal and spectral modelling as detailed in \S \ref{s:timing} and \S \ref{s:spec}.
 
\section{Timing Analysis and Results}

\label{s:timing}

We generate the light curves by combining data from {\it LAXPC10} and {\it LAXPC20}
with $1$ s binning for generating CCDs. 
CCDs are used to understand the hardness ratio variation and variability class of the source. 
Following this, we generate $1$ ms binned light curves for generating the power spectra. The modelling of power spectra and the obtained results are presented in \S \ref{s:temp_pro}.

\subsection{Color-Color Diagram}

\begin{figure}
\begin{center}
\includegraphics[height=8.5cm, width=9cm]{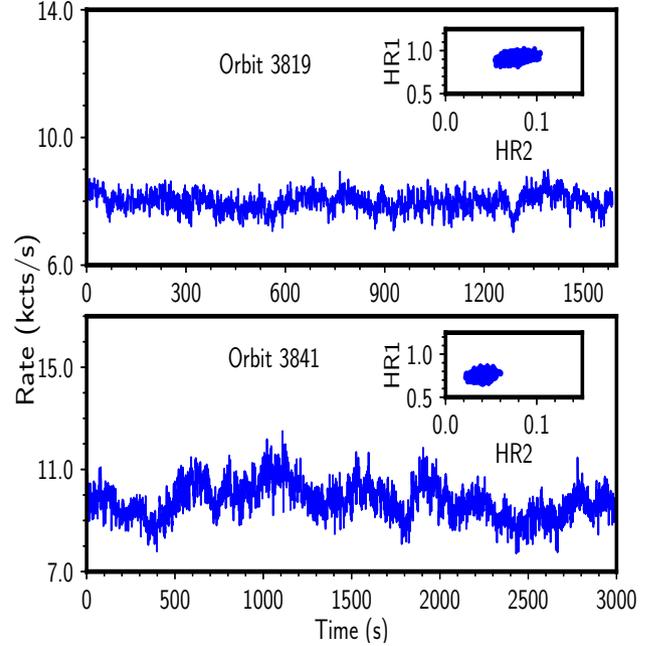} 
\end{center}

\caption{Dead-time corrected light curve in the energy
range $3 - 60$~keV for combined observations with {\it LAXPC10} and
{\it LAXPC20}. The color-color diagram is shown as inset of each
panel. The top panel corresponds to the observation during Orbit
3819 where HFQPO is detected and the bottom
panel corresponds to the observation during Orbit 3841, where
HFQPO is absent. See text for details.}
\label{fig:LCnCCD}
\end{figure}

As mentioned before, we use light curves with 1 s binning for generating CCDs.
The CCD is the variation of ${\rm HR}_1=B/A$ versus ${\rm HR}_2=C/A$, where $A$, $B$ and $C$ are the count rates in $3 - 6$ keV, $6 - 15$ keV and 
$15 - 60$ keV energy bands, respectively. The background subtracted and dead-time corrected (see equation \ref{eq:rate_in}) light curves in the energy band $3 - 60$ keV for the {\it LAXPC} observations on MJD 57551.04
(Orbit 3819) and MJD 57552.56 (Orbit 3841) are shown in
Figure \ref{fig:LCnCCD}. The background count rate is in the range of 100 to 140 counts/sec. The CCDs
are provided as inset in each panel which indicate that the maximum values of HR$_1$ and HR$_2$ for both the observations are less than $1.1$ and $0.12$, respectively. Comparing the CCDs and X-ray variability with
{\it RXTE} observations \citep{Belloni-etal00}, we classify that the source is in its $\delta$ class during our observational campaign.
However, both HR$_1$ and HR$_2$ values are seen to decrease as the source count rate increases whereas the reverse trend is observed towards the end (see Fig. \ref{fig:allqpos} and Table \ref{tab:laxpc_grs_qpo}). These findings indicate the signature of soft state (see Table \ref{tab:bbspec_pheno} for spectral parameters). In addition, we do not find HFQPOs in those observations where mean values of HR$_1$ $\le 0.76$ and HR$_2$ $\le 0.04$ (see Table \ref{tab:laxpc_grs_qpo} and \S \ref{s:temp_pro}). In all the remaining observations, we detect HFQPOs.
 
\subsection{Temporal Properties}
\label{s:temp_pro}
We generate power spectrum for each {\it LAXPC} observation considering a Nyquist frequency of 500 Hz in order to search for the presence of HFQPOs. We use intervals of $32768$ bins and compute the power spectrum for each interval, which is then averaged to obtain the final power spectrum. A geometric binning factor of 1.03 is used for representing the power spectrum in frequency space. After generating the Leahy power spectrum \citep{Leahy1983}, we carry out dead-time ($\tau_{\rm d}$) corrected Poisson noise subtraction following \cite{Agrawal-etal2018,Sreehari-etal2019}. Dead-time corresponds to the time interval between the successive photon detection by the detector. 
This effectively affects the poisson noise level yielding the reduction of actual rms in the power spectrum.
{\it LAXPC} has a dead-time of 42.5~$\mu$s \citep{Yadav-etal2016}. Following \cite{VanderKlis1989}, we calculate the incident count rate ($r_{\rm in}$) as

\begin{equation}
r_{\rm in} = \frac{\displaystyle r_{\rm det}}{\displaystyle (1 - \tau_{\rm d} \: r_{\rm det}/N)},
\label{eq:rate_in}
\end{equation}

where $r_{\rm det}$ refers to the detected count rate and $N$ is the number of proportional counter units used.

%

Thereafter, following \citet[and references therein]{Zhang1995}, we compute the dead-time affected Poisson noise power ($P_{\rm n}$). Besides affecting the noise level, dead-time also modulates the source rms. So, after subtracting $P_{\rm n}$ level from the power spectrum, we correct for the dead-time effects on rms amplitude by scaling it using the relation given by,

\begin{equation}
{\rm rms}_{\rm in} = {\rm rms}_{\rm det} \: (1 + \tau _{\rm d} \: {\rm r}_{\rm in}/N) = \frac{{\rm rms}_{\rm det}}{(1 - \tau _{\rm d} \: {\rm r}_{\rm det}/N)},
\label{eq:rms_dead1}
\end{equation} 
where ${\rm rms}_{\rm in}$ is the dead-time corrected rms and ${\rm rms}_{\rm det}$ is the rms detected by the instrument \citep[see][for details]{Bachetti2015, Sreehari-etal2019}.

We model the power density spectrum (PDS) using a combination of {\it constant} and  {\it Lorentzian} features, and illustrate it in frequency (Hz)  versus ${\rm rms}^2/$Hz plane.
The {\it Lorentzian} has three parameters namely centroid frequency ($\nu$), full width at half maximum ($FWHM$) and normalization (norm). Here, a {\it Lorentzian} feature is adopted to define QPO, based on the values of its Quality factor (Q = $\nu/FWHM$), 
significance and rms \citep[see][and references therein]{Sreehari-etal2019}. 
The QPO rms is computed by finding the square root of the definite integral of the {\it Lorentzian} representing the QPO in Frequency-power space. The QPO significance is computed as the ratio of {\it Lorentzian} normalization to its negative error
\citep[see][and references therein]{Alam2014,Sreehari-etal2019}.

\begin{figure}
\begin{center}
\includegraphics[height=9.0cm, width=7.0cm,angle=-90]{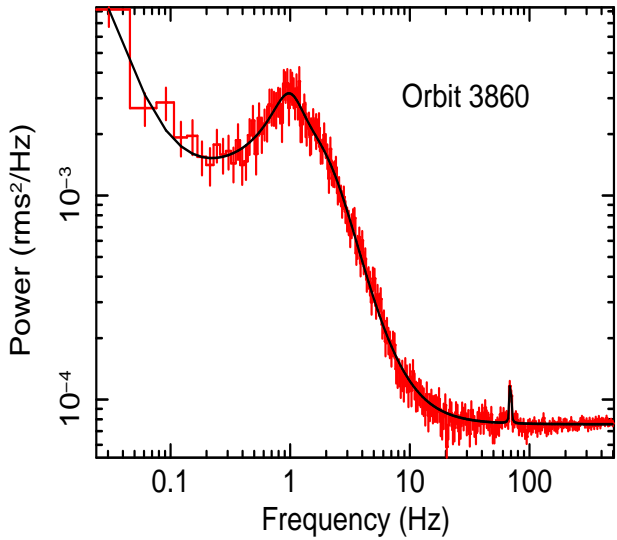} 
\end{center}

\caption{Power spectrum for the observation of Orbit 3860 from combined {\it LAXPC10} and {\it LAXPC20} data in the broad frequency range from $0.06 -500$ Hz. PDS is modelled with multiple \texttt{Lorentzians} and a \texttt{constant} component.}
\label{fig:pdsfull}
\end{figure}

Figure \ref{fig:pdsfull} presents the PDS from combined {\it LAXPC10} and {\it LAXPC20} data corresponding to the Orbit 3860 in the frequency range from $0.06$ Hz to $500$ Hz. The PDS is modelled with multiple \texttt{Lorentzians} and a \texttt{constant} that yields $\chi_{\rm red} ^2 = 135.2/208=0.65$. Since the detected HFQPOs are seen to lie within a narrow range of frequencies, for the purpose of representation, we model the power spectra in the frequency range $20 - 200$ Hz for all the observations.

\begin{figure}
\begin{center}
\includegraphics[height=12cm, width=8.2cm]{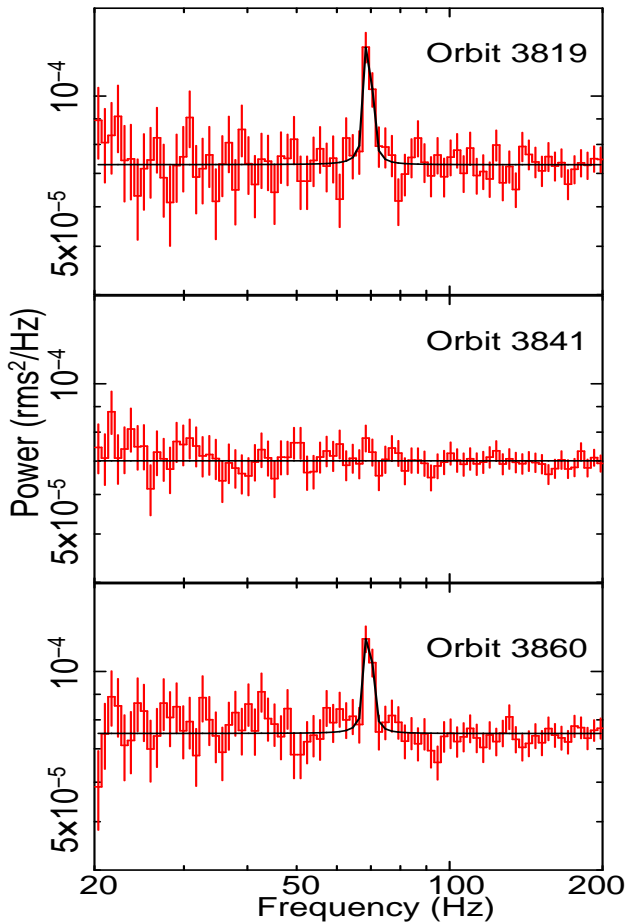} 
\end{center}

\caption{Combined power spectra of {\it LAXPC10} and {\it LAXPC20} in $3 - 25$~keV energy band, for three orbits of {\it AstroSat} observations. HFQPOs are detected during Orbits 3819 and 3860, whereas no such detection is seen during Orbit 3841 (middle panel). See text for details.}
\label{fig:pds1}
\end{figure}

Following the above consideration,  
we fit the PDS in the said frequency range ($i.e.$, $20-200$ Hz) with a {\it constant}
and a {\it Lorentzian}. The modelled PDS for three {\it LAXPC} observations are shown in Figure \ref{fig:pds1}. 
The top panel (Orbit 3819) and the bottom panel (Orbit 3860) indicate detection of HFQPOs around 69 Hz.
The middle panel representing the observation during Orbit 3841 does not indicate any HFQPO feature.
Following \cite{Belloni2001}, we calculate the normalization of the \texttt{Lorentzian} feature in this observation (Orbit 3841) by freezing the \texttt{Lorentzian} centroid at 68.83 Hz and width at 1.4 Hz from the previous observation. The corresponding upper limit on significance is only 0.61.
During our {\it AstroSat} campaign, $18$ more such HFQPO signatures are detected having 
frequencies in the range of $67.96 - 70.62$ Hz. The model fitted parameters of these HFQPOs like centroid frequency, FWHM, significance and rms along with the HR variations are presented in Table \ref{tab:laxpc_grs_qpo}. These detections are consistent with the \textit{RXTE} observation of variable HFQPOs ($63.5-71.3$ Hz), with an average of $67.3 \pm 2.0$ Hz \citep{Morgan1997,Belloni-Altamirano13}. 

\begin{figure*}
\begin{center}
\includegraphics[height=7.4cm, width=8.5cm]{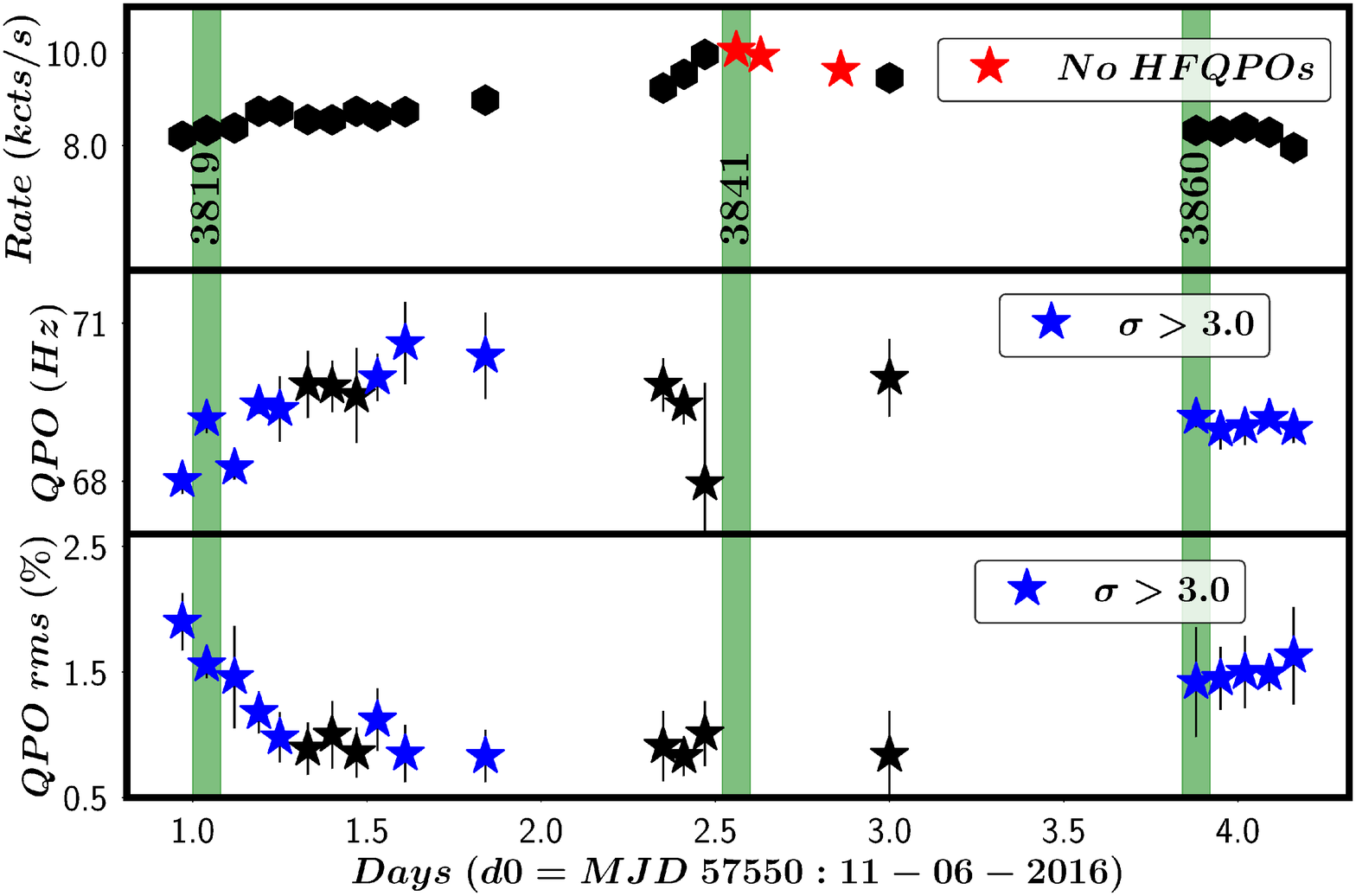} 
\includegraphics[height=7.4cm, width=8.5cm]{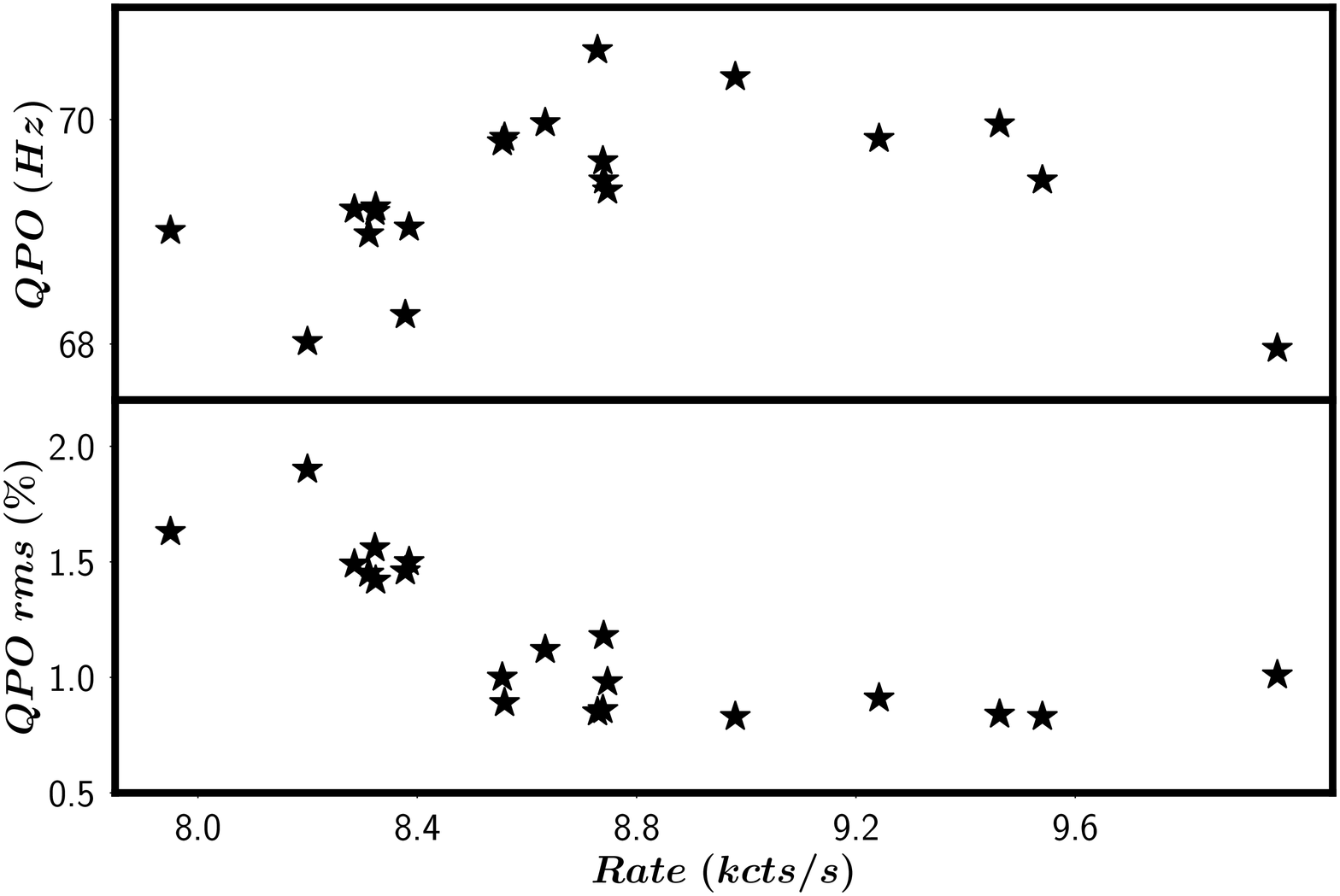}
\end{center}

\caption{LHS: Top panel depicts the dead-time corrected source count rate for each observation. The red asterisks indicate observations where HFQPO is absent. The middle panel shows the frequency of all detected HFQPOs and the ones above 3 $\sigma$ significance are presented in blue colour. The bottom panel shows the corresponding HFQPO rms. The vertical green shades represent the observations for which the power spectra are presented in Figure \ref{fig:pds1}. RHS: The variation of QPO frequency with count rate is shown in the top panel and the dependence of QPO rms on count rate is presented in the bottom panel.}
\label{fig:allqpos}
\end{figure*}

In Figure \ref{fig:allqpos} (left), we present the time evolution of the source count rate (top panel), HFQPO frequency (middle panel) and QPO rms (bottom panel). The variation of count rate within the soft state spanned over a few days is also to be noted. In the top panel, the red stars indicate the observations in which HFQPOs are absent. In the middle and bottom panels, all the HFQPO detections are presented and the blue stars indicate the HFQPOs with significance above 3. Figure \ref{fig:allqpos} (right) shows the correlation of QPO frequency and QPO rms with count rate.
It is evident that as the source count rate increases and HR value decreases (see Table \ref{tab:laxpc_grs_qpo}), the HFQPO rms reduces and eventually disappears. The vertical green bars indicate the observations during Orbits 3819, 3841 and 3860 that we present in Figure \ref{fig:pds1}. From Table \ref{tab:laxpc_grs_qpo}, we find that the percentage rms of HFQPO lies in the range of 0.83\% to 1.90\%. The dead-time correction factor for rms is obtained in the range 1.17 to 1.21. The HFQPO is weak or below the significance level of detection, when the source count rate peaks. 

\begin{table*}
\caption[HFQPO parameters from LAXPC observations of GRS 1915$+$105]{Details of HFQPO parameters from {\it LAXPC} observations of GRS 1915$+$105 in $3 - 25$~keV energy band. These results are obtained from the observational data of {\it LAXPC10} and {\it LAXPC20} during $\delta$ class variability of the source. Errors corresponding to 68\% confidence are quoted for each parameter. See text for details.}
\resizebox{\textwidth}{!}{\begin{tabular}{|c|c|c|c|c|c|c|c|c|c|c|c|c|c|}
\hline
	MJD   &     Orbit &    Exposure &  $r_{\rm det}$ (cts/s) & $r_{\rm in}$ (cts/s) & QPO (Hz)  &  FWHM (Hz)   & Significance ($\sigma$) & rms (\%)  & HR$_1$ &  HR$_2$ &  $\chi ^2$/dof  \\
\hline                                                                                             

57550.97  &  3818       &    1982 &       6835  & 7996     &     68.02$_{-0.24}^{+0.29}$   &     2.82$_{-0.64}^{+0.54}$     &   7.75      &    1.90 $\pm$ 0.23 & 0.92 & 0.08 &    48/74 \\  
{\bf 57551.04}  &  {\bf 3819} &    1590 &  6919  & 8112   &      69.18$_{-0.27}^{+0.14}$   &     1.49 fixed     &   6.94   &    1.56 $\pm$ 0.11 & 0.91 & 0.08 &   43/75 \\  
57551.12  &  3820       &    1030 &       6954  & 8159   &      68.26$_{-0.28}^{+0.17}$   &     2.33$_{-1.11}^{+1.12}$      &   3.45    &    1.46 $\pm$ 0.41& 0.91 & 0.08 &   34/75 \\
57551.19  &  3821       &    0791 &       7214  & 8520   &      69.46$_{-0.15}^{+0.21}$   &     1.27$_{-0.21}^{+0.22}$      &   4.17 &      1.18 $\pm$ 0.17& 0.88 & 0.07 &   50/75     \\  
57551.25  &  3822       &    1215 &       7223  & 8533   &      69.37$_{-0.70}^{+0.55}$   &     1.72$_{-0.42}^{+0.48}$      &  3.19 &    0.98 $\pm$ 0.20& 0.86 & 0.06 &   32/75 \\
57551.33  &  3823       &    1638 &       7093  & 8352   &      69.84$_{-0.36}^{+0.92}$   &     1.72 fixed     &   2.06     &    0.89 $\pm$ 0.21& 0.85 & 0.06 &   43/75 \\ 
57551.40  &  3824       &    2063 &       7095  & 8355   &      69.78$_{-0.26}^{+0.76}$   &     1.67$_{-0.41}^{+0.50}$     &   2.16   &    1.00 $\pm$ 0.27& 0.83 & 0.06 &   32/75 \\ 
57551.47  &  3825       &    2516 &       7162  & 8448   &      69.63$_{-0.34}^{+1.44}$   &     1.84$_{-0.53}^{+0.67}$     &   2.94    &    0.86 $\pm$ 0.20& 0.84 & 0.06 &    39/75 \\
57551.53  &  3826       &    2941 &       7153  & 8435   &      69.97$_{-0.32}^{+0.58}$   &     2.73$_{-0.88}^{+1.10}$     &   3.86	    &    1.12 $\pm$ 0.25&  0.84 & 0.06 &  46/75 \\
57551.61  &  3827       &    3364 &       7224  & 8534   &  	70.62$_{-0.93}^{+0.63}$   &     2.31$_{-1.08}^{+1.18}$     &   4.26    &    0.85 $\pm$ 0.23& 0.83 & 0.06 &    50/75 \\   
57551.84  &  3830       &    2953 &       7405  & 8787   &      70.38$_{-1.01}^{+0.64}$   &     2.04$_{-1.00}^{+0.94}$     &   3.84    &    0.83 $\pm$ 0.21& 0.82 & 0.05 &   34/75 \\
57552.35  &  3838       &    1116 &       7586  & 9045   &      69.83$_{-0.27}^{+0.95}$   &     1.18 fixed     &  1.63    &    0.91 $\pm$ 0.28& 0.77 & 0.04 &   42/75 \\ 
57552.41  &  3839       &    2180 &       7793  & 9339   &      69.46$_{-0.35}^{+0.41}$   &     1.46 fixed    &   2.49    &    0.83 $\pm$ 0.16& 0.77 & 0.04 &   26/75 \\   
57552.47  &  3840       &    2603 &       8089  & 9768   &      67.96$_{-1.84}^{+1.99}$   &     5.37$_{-1.85}^{+2.27}$     &   2.95     &    1.01 $\pm$ 0.26 & 	0.76 & 0.04 &   47/77 \\
{\bf 57552.56}  &  {\bf 3841} &    3027 &       8155  & 9865   &      -	      	   &     -     	        &    -       &    -              &  0.75 & 0.04 &  39/77     \\ 
57552.63  &  3843       &    2154 &       8076  & 9749   &      -	      	       &     -     	        &    -       &    -              &  0.75 & 0.04 &  41/77     \\
57552.86  &  3845       &    2402 &       7867  & 9446   &      -	      	       &     -     	        &    -       &    -              &  0.76 & 0.04 &  38/77     \\
57553.00  &  3848       &    1939 &       7738  & 9260   &      69.96$_{-0.50}^{+1.08}$   &     2.57$_{-0.77}^{+2.32}$     &   2.49    &    0.84 $\pm$ 0.35& 0.77 & 0.04 &   36/75 \\
{\bf 57553.88}  &  {\bf 3860} &    2381 &       6813  & 7966   &69.22$_{-0.24}^{+0.17}$   &     2.53$_{-1.63}^{+1.18}$     &   3.74    &    1.42 $\pm$ 0.44& 0.88 & 0.07 &   45/75 \\ 
57553.95  &  3862       &    2042 &       6918  & 8110   &      68.98$_{-0.43}^{+0.33}$   &     3.78$_{-1.12}^{+1.15}$     &   5.36   &    1.45 $\pm$ 0.25& 0.88 & 0.07 &   34/75 \\ 
57554.02  &  3863       &    1755 &       6965  & 8176   &      69.03$_{-0.41}^{+0.29}$   &     3.35$_{-1.23}^{+1.04}$     &   5.23   &    1.50 $\pm$ 0.29& 0.88 & 0.07 &   38/75 \\ 
57554.09  &  3864       &    1170 &       6888  & 8070   &      69.20$_{-0.14}^{+0.15}$   &     2.00$_{-0.26}^{+0.27}$     &   7.43    &    1.49 $\pm$ 0.14&  0.89 & 0.07 &  41/75 \\
57554.16  &  3865       &    1156 &       6638  & 7728   &      69.01$_{-0.33}^{+0.24}$   &     2.55$_{-1.33}^{+0.87}$     &   4.61    &    1.63 $\pm$ 0.39& 0.90 & 0.08 &   36/75 \\ 
\hline
\end{tabular}}

\label{tab:laxpc_grs_qpo}
\end{table*}

The $\chi_{\rm red}^2$ values for power spectral modelling are provided in Table \ref{tab:laxpc_grs_qpo}. Modelling the PDS (Orbit 3860) in the frequency range $20 - 200$~Hz with only a constant yields a $\chi_{\rm red}^2$ of $97.68/78=1.25$. However, the presence of excess power above the constant level around 69 Hz indicates the presence of a QPO. This feature is further modelled with a \texttt{Lorentzian} apart from the constant resulting in an overall $\chi_{\rm red}^2$ of $45/75=0.6$. 
The decrease of $\chi_{\rm red}^2$ from 1.25 to 0.6 suggests that HFQPO feature is significant and requires modelling.
Moreover, we detect the \texttt{Lorentzian} feature at ~69 Hz in several observations which indicates that it is not a random increase in amplitude and hence we argue that it is not a case of over-fitting.

\begin{figure}
\begin{center}
\includegraphics[height=9.7cm, width=8.5cm]{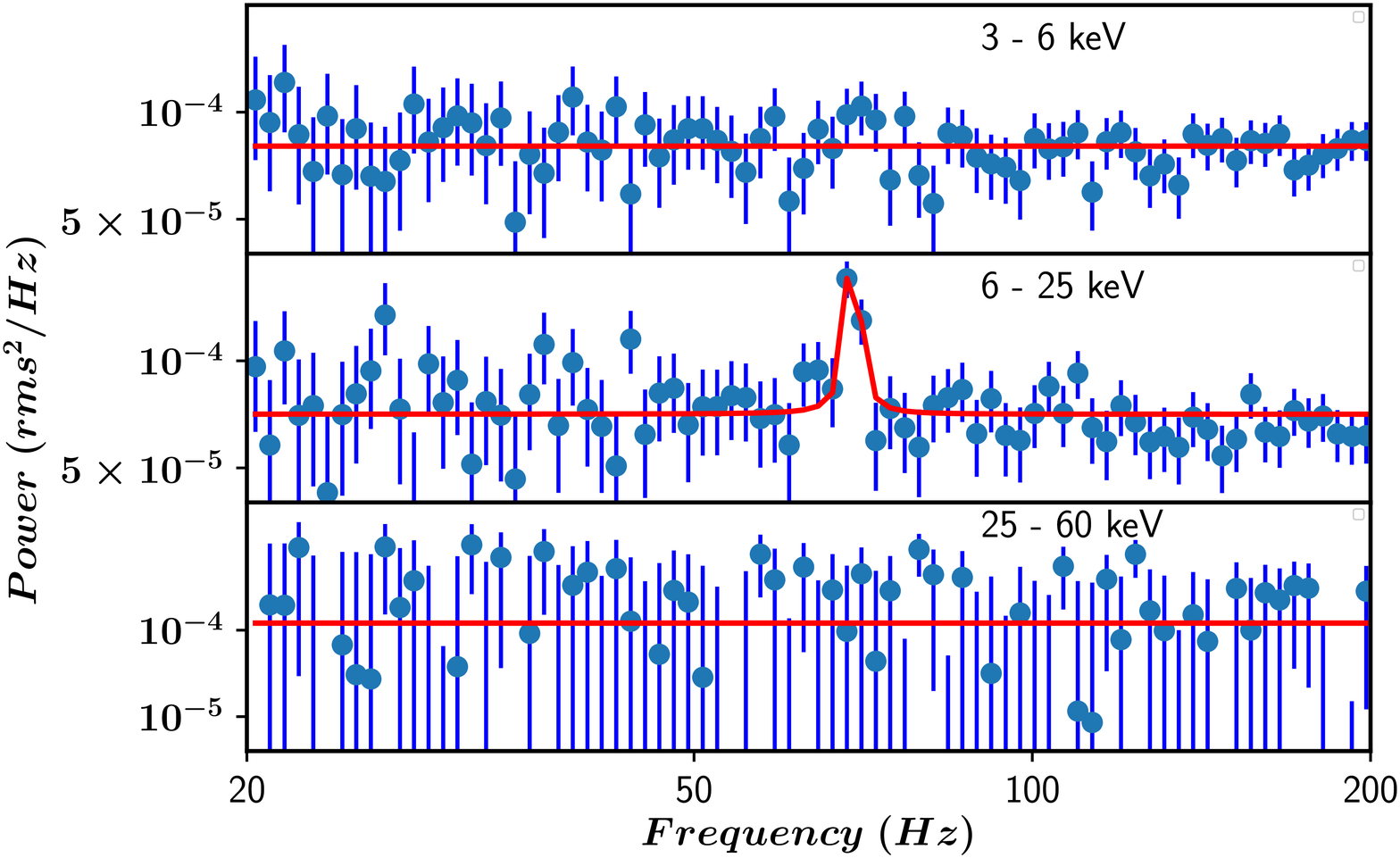} 
\end{center}
\caption{The energy dependent power spectra for the observation corresponding to Orbit 3819 are shown. Top panel indicates that there is no signature of QPO in the $3 - 6$~keV energy band. The middle panel corresponding to $6 - 25$ keV energy band shows significant detection of HFQPO. Bottom panel is for $25 - 60$~keV energy band where power spectrum is noise dominated without any detection of QPO. See text for details.}
\label{fig:Endep3819}
\end{figure}

Since HFQPOs are detected in several observations, we intend to carry out energy dependent behaviour of these oscillations. For this purpose, we generate PDS from light curves in the energy range of $3 - 6$
keV, $6 - 25$ keV and $25 - 60$ keV and model them in the frequency range of $20 - 200$ Hz.
Figure \ref{fig:Endep3819} shows the energy dependent PDS for MJD 57551.04 (Orbit 3819).
It is evident that the HFQPO is present only in the $6 - 25$ keV energy band (middle panel) and not in the lower ($3 - 6$ keV) and higher ($25 - 60$ keV) energy bands. We find that the upper limit of the significance in the energy range $3 - 6$ keV and $25-60$ keV are $1.69$ and $0.44$, respectively. The strong HFQPO feature (Q-factor~$\sim~46$, significance of $6.45~ \sigma$ and rms of $2.15 \pm 0.23$\%) in the $6 - 25$~keV energy band is found at $68.75^{+0.12}_{-0.12}$~Hz. All {\it LAXPC} observations of GRS 1915$+$105 have similar power spectral features (non-detection of HFQPO signals) in the $3 - 6$~keV and $25 - 60$~keV band. However, the HFQPO strength (rms) and significance varies over the observations as can be seen from Table \ref{tab:laxpc_grs_qpo}.
The absence of HFQPO feature in $3 - 6$~keV energy band indicates that the QPOs are likely to be the result of an oscillating corona located in the vicinity of the black hole and it emits in $6 - 25$~keV energy band.
In order to have better clarity on the energy dependence of HFQPOs and understand the emission processes, we carry out the broadband energy spectral modelling of {\it AstroSat} data of GRS~1915$+$105 and present the results in \S \ref{s:spec}.

\begin{figure*}
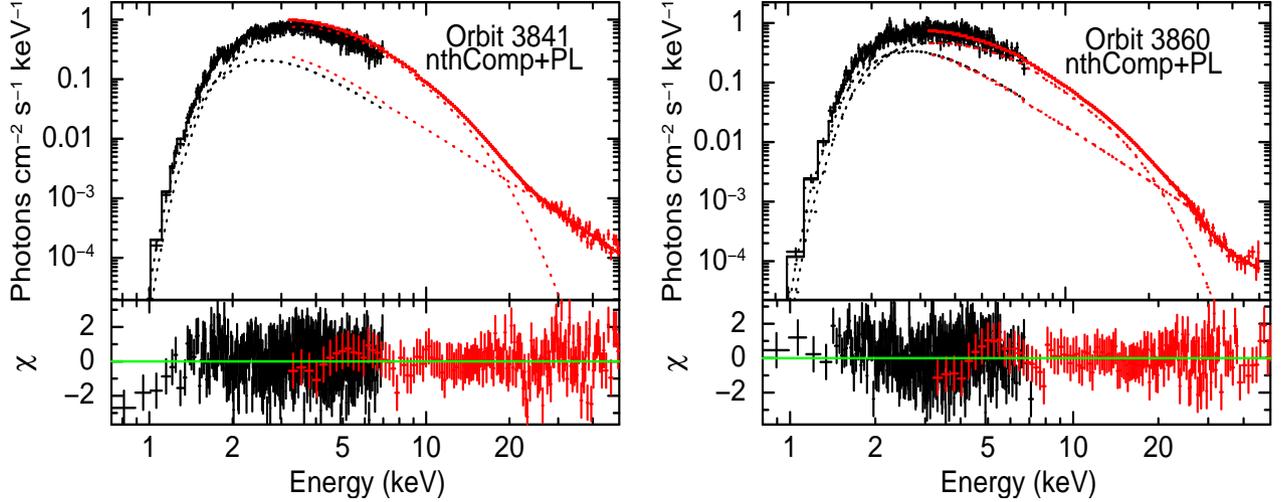

	\begin{center}
		\includegraphics[trim=0 0 0.25mm 0, clip = true, width=0.39\textwidth, height=8.5cm, angle=-90]{fig6a.eps}
		\includegraphics[trim=0 0 0.25mm 0, clip = true, width=0.39\textwidth, height=8.5cm, angle=-90]{fig6b.eps}
		
	\end{center}
	\caption{The best fit unfolded energy spectra ($0.7 - 50$~keV) of GRS 1915$+$105 observed 
		on MJD 57552.56 (Orbit 3841) and MJD 57553.88 (Orbit 3860) with {\it AstroSat}. The spectra are modelled with \texttt{Tbabs(nthComp $+$ powerlaw)}. The bottom panel of each spectrum shows residuals in units of 
		$\sigma$. See text for details.}
	\label{fig:spec_bb_pheno}
\end{figure*}

\begin{table*}
	\caption{Broadband spectral modelling parameters from the model \texttt{Tbabs(nthComp $+$ powerlaw)}. Observations in which HFQPOs are absent are tabulated separately. Errors are computed with 90\% confidence.}
	\centering 
	\begin{tabular}{@{}lcccccccc} \hline \hline
		MJD (Orbit)& $kT_{\rm bb}$ (keV) &  $kT_{e}$ (keV) & $\Gamma _{\rm nth}$ & norm$_{\rm nth}$ & $\Gamma_{\rm PL}$ & norm$_{\rm PL}$ & Flux (ergs/cm$^2$/s) &$\chi ^2/{\rm dof}$\\
		\hline
		\multicolumn{6}{|c|}{Detection of HFQPOs} \\ 
		\hline

		57551.33 (3823)  & 0.3 & 2.25 $\pm$ 0.08 & 1.79 $\pm$ 0.07 & 8.1 & 2.94 $\pm$ 0.12 & 15.8 & 3.09$~\times ~10^{-8}$& 440/415 = 1.06\\
		57551.53 (3826)  & 0.19 $\pm$ 0.02 & 2.32 $\pm$ 0.03 & 1.76 $\pm$ 0.01 & 8.6 & 3.04 $\pm$ 0.01 &17.6& 3.11$~\times ~10^{-8}$& 655/551 = 1.18 \\
		57551.84 (3830)  & 0.1 & 2.23 $\pm$ 0.02 & 1.73 $\pm$ 0.03 & 8.9 & 3.04 $\pm$ 0.04 &15.5&3.22$~\times ~10^{-8}$& 737/628 = 1.17 \\
		57552.41 (3839)$\dagger$  & 0.1 & 2.07 $\pm$ 0.18 & 2.24 $\pm$ 0.86 & 3.9 & 3.28 $\pm$ 0.03 &34.9&3.34$~\times ~10^{-8}$& 441/339 = 1.30\\  
		57553.88 (3860)  & 0.1 & 2.43 $\pm$ 0.13 & 1.74 $\pm$ 0.06 & 7.0 & 3.09 $\pm$ 0.04 &25.5&3.07$~\times ~10^{-8}$& 615/547 = 1.12	\\
		
		\hline
		\multicolumn{6}{|c|}{Non-detection of HFQPOs} \\ 
		\hline 

		57552.56 (3841) & 0.1 & 2.09 $\pm$ 0.15  & 2.45 $\pm$ 0.84 & 3.01 & 3.28 $\pm$ 0.02 &30&3.55$~\times ~10^{-8}$& 729/636 = 1.14\\ 
		\hline \hline
		$\dagger:$ Weakest HFQPO&&&&&&&&
	\end{tabular}
	\label{tab:bbspec_pheno}
	
\end{table*}


\section{Spectral Analysis and Results}
\label{s:spec}

To carry out the spectral analysis, we use simultaneous data from both {\it SXT} and {\it LAXPC}. 
The latest calibration files are provided by {\it AstroSat} mission 
team\footnote{\url{http://astrosat-ssc.iucaa.in/?q=laxpcData}}. A systematic of 2\% is
added per spectral bin as suggested by \cite{Antia-etal17, Leahy2019}. 

{\it SXT} spectra are extracted in the $0.7 - 7$~keV band, whereas dead-time corrected {\it LAXPC} spectra \citep{Antia-etal17} are extracted in the energy range of $3 - 50$ keV \citep[see][for details]{Agrawal-etal2018,Sreehari-etal2019}. As 
there are only a few good quality data available from {\it SXT} in the 
database\footnote{\url{https://astrobrowse.issdc.gov.in/astro\_archive/archive/Home.jsp}}, 
we are able to generate broadband spectra only for those observations. Modelling of the {\it SXT} spectra indicates an nH column density (in $10^{22}$ atoms/cm$^2$) in the range 5.93 $\pm$ 1.01 to 6.07 $\pm$ 1.19. We model the energy spectra with fixed nH of $6 \times 10^{22}$ atoms/cm$^2$ and the obtained spectral parameters are quoted in Table \ref{tab:bbspec_pheno}. The change in parameters due to the variation of nH is within the error bars of the values obtained by freezing nH at $6 \times 10^{22}$ atoms/cm$^2$. Gain correction is applied using \texttt{gain fit} routine of \textsc{XSPEC} on all the {\it SXT} spectra to account for the instrumental features peaks at $1.8$ and $2.2$ keV.
Initially, we model the broadband spectra with phenomenological model combination  
\texttt{Tbabs(diskbb $+$ powerlaw $\times$ smedge)} that yields a disc 
temperature of $2.57 \pm 0.05$~keV, photon index of $3.11 \pm 0.06$ for the observation 
on MJD 57551.33 (Orbit 3823) with a $\chi^2/{\rm dof} = 438/418 = 1.04$. Using \texttt{cflux} model, we estimate the disc contribution which is found to be 69.7\%  of the total flux. This indicates that the source is in thermally dominated soft state. This also corroborates the CCDs representing the light curve ($i.e.$, $\delta$ class) as a softer state of GRS 1915$+$105.

In order to understand the physical processes leading to the emission
from the source, we fit the broadband energy spectrum of
Orbit 3823 with the thermal Comptonization model (\texttt{nthComp} in \textsc{XSPEC}; \cite{Zdziarski_etal1996}). The fit resulted with $\chi_{\rm red}^2=\chi^2/{\rm dof}=1534/420=3.65$ as the higher 
energy part of the spectrum (above 30 keV) is not fitted with \texttt{nthComp} model. Upon inclusion of an 
additional power-law component, the combined models provide acceptable fit with $\chi_{\rm red}^2=\chi^2/{\rm dof}=440/415=1.06$. Hereafter, we carry out the broadband spectral modelling with the combined models defined as \texttt{Tbabs(nthComp + powerlaw)}.
The model parameters of \texttt{nthComp} are electron temperature 
($kT_e$) equal to $2.25 \pm 0.08$~keV, seed-photon temperature ($kT_{\rm bb}$) of $\sim 0.30$~keV and photon-index ($\Gamma_{\rm nth}$) of $1.79 \pm 0.07$. An additional \texttt{powerlaw} 
component with a photon-index ($\Gamma_{\rm PL}$) of $2.94 \pm 0.12$ is required to fit the 
high energy part of the spectrum (above $\sim 25$ keV). Following this 
approach, we model the broadband spectral data for all the available observations irrespective of the presence of HFQPOs (see Table \ref{tab:bbspec_pheno}).

In the left panel of Figure \ref{fig:spec_bb_pheno}, we depict the unfolded spectrum
of an observation (MJD 57552.56, Orbit 3841) for which HFQPO feature is not seen in the PDS
(middle panel of Figure \ref{fig:pds1}). 
The fitted parameters are obtained as $kT_e$ $\sim 2.09$~keV 
and $\Gamma_{\rm nth} =2.45 \pm 0.84$, respectively. The \texttt{powerlaw} index in this case 
is obtained as $\Gamma_{\rm PL} \sim 3.28$. 
Further, we model the broadband energy spectrum for the observation on MJD 57553.88 
(Orbit 3860) which shows a significant ($3.74\sigma$) detection of HFQPO in the PDS (bottom panel of Figure \ref{fig:pds1}). 
In the right panel of Figure \ref{fig:spec_bb_pheno}, we present the unfolded spectrum modelled 
with \texttt{Tbabs(nthComp $\times$ smedge $+$ powerlaw)}. The \texttt{nthComp} 
component yields $\Gamma_{\rm nth} = 1.74 \pm 0.06$ and $kT_e \sim 2.43$ keV. 
The \texttt{powerlaw} component is seen to be strong with a steep photon index of $\Gamma_{\rm PL} =3.09 \pm 0.04$. The additional \texttt{smedge} component which is not required for orbit $3841$,  is however used to model the absorption feature around 7 keV. 
The model fitted parameters for all the broadband observations are presented in Table \ref{tab:bbspec_pheno}.
It may also be noted that the energy spectral parameters for Orbit 3839, which has the detection of the weakest HFQPO (rms~$\sim 0.83$\% and significance~$\sim 2.49$) corresponds to a weak \texttt{nthComp} (${\rm norm}_{\rm nth} \sim 3.9$) and a strong \texttt{powerlaw} (${\rm norm}_{\rm PL}~\sim~34.9$) contribution. We discuss the implications of these results in \S \ref{s:disc_grs}.

\section{Estimation of Mass, Spin and Accretion rate}

\label{s:kerr}

\begin{figure*}
	\begin{center}
		\includegraphics[trim=0 0 0.25mm 0, clip = true, width=0.39\textwidth, height=8.5cm, angle=-90]{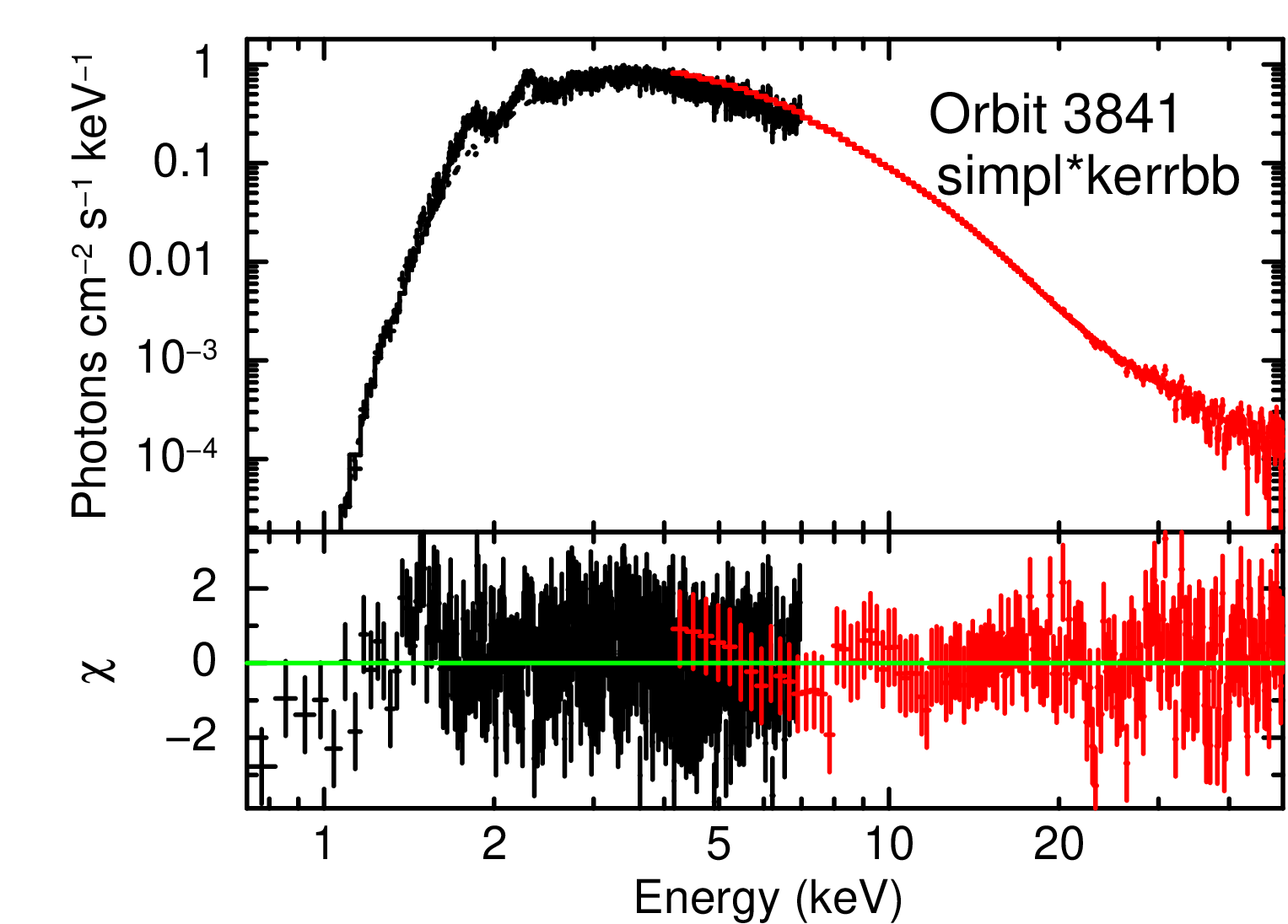}
		\includegraphics[trim=0 0 0.25mm 0, clip = true, width=0.39\textwidth, height=8.5cm, angle=-90]{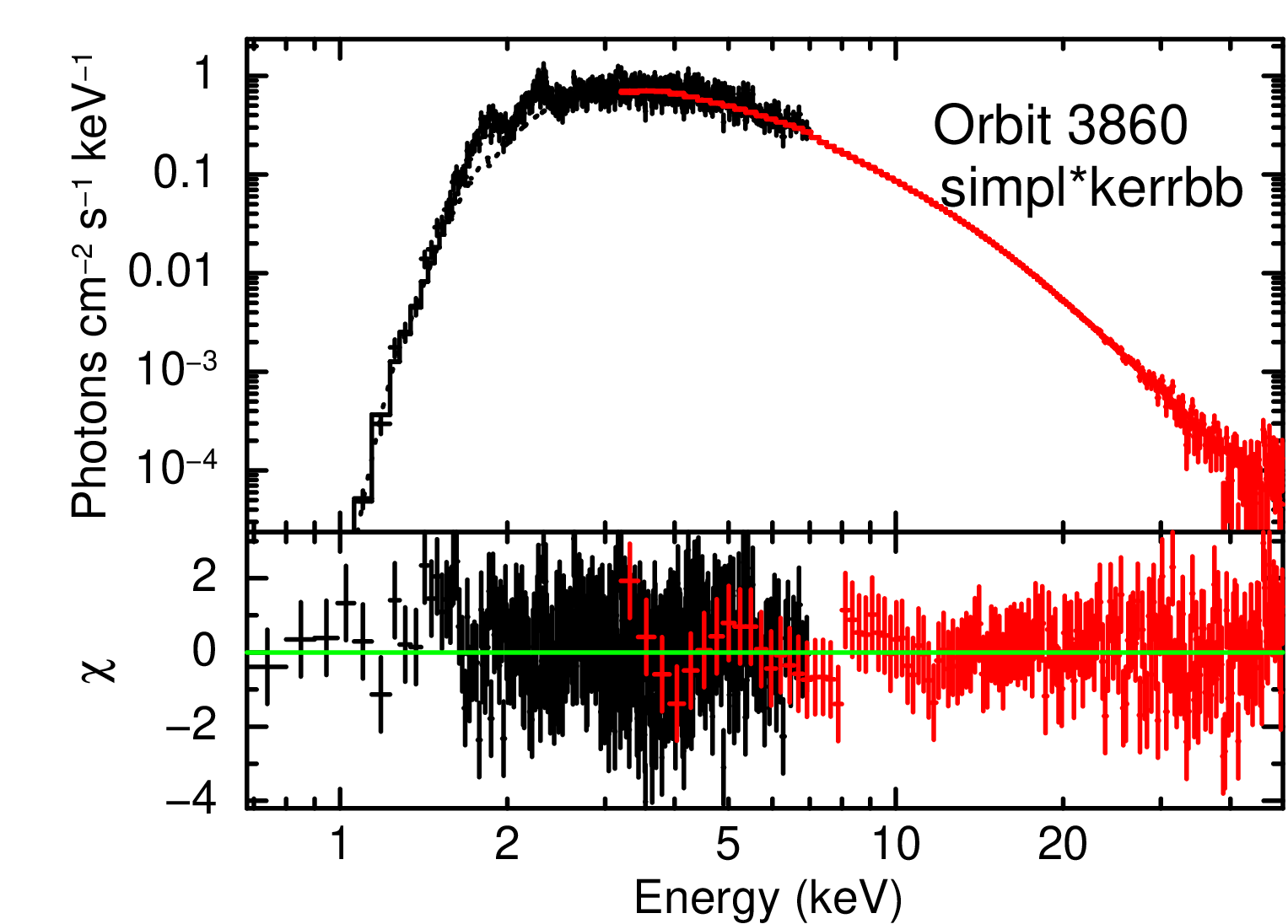}
	\end{center}
	\caption{The best fit unfolded energy spectra ($0.7 - 50$ keV) of GRS 1915+105 
		observed on MJD 57552.56 (Orbit 3841) and MJD 57553.88 (Orbit 3860) with {\it AstroSat}. The spectra are modelled with 
		\texttt{Tbabs(simpl*kerrbb $\times$ smedge)constant}. 
		The bottom panel of each figure shows residuals in units of $\sigma$.}
	\label{fig:kerrbbspec}
\end{figure*}

\begin{table*}
	\caption{Broadband spectral modelling parameters from \texttt{Tbabs(simpl * kerbb $\times$ smedge)constant}. Errors are computed with 90\% confidence.}
	\centering 
	\resizebox{\textwidth}{1.3cm}{
		\begin{tabular}{@{}lcccccccc} \hline \hline
			MJD (Orbit)& $\Gamma$ &  Frac. Scattered (${\rm F}_{\rm sca}$) & $\dot{M}$ (10$^{18}$~g/s) & Mass (M$_{\odot}$) & $a_{\rm k}$ &E$_{\rm smedge}$ (keV)& $\chi ^2/{\rm dof}$\\
			\hline
			57551.53 (3826) & 3.92 $\pm$ 0.01 & 0.13 $\pm$ 0.01  & 7.65 $\pm$ 0.11  & 12.66 $\pm$ 0.22 & 0.996 $\pm$ 0.001 &6.41 $\pm$ 0.42 &(657/564=1.16)\\

			57551.84 (3830) & 3.83 $\pm$ 0.01 & 0.09 $\pm$ 0.01  & 7.73 $\pm$ 0.07  & 12.75 $\pm$ 0.16 & 0.996 $\pm$ 0.001 &6.44 $\pm$ 0.40 &(733/629=1.16)\\

			57552.56 (3841) & 3.09 $\pm$ 0.01 & 0.05 $\pm$ 0.01  & 7.89 $\pm$ 0.07  & 12.92 $\pm$ 0.17 & 0.991 $\pm$ 0.001 &7.88 $\pm$ 0.86 &(717/639=1.12)\\
			
			57553.88 (3860)  & 4.80 $\pm$ 0.01 & 0.55 $\pm$ 0.01 & 7.05 $\pm$ 0.10 & 12.85 $\pm$ 0.24 & 0.995 $\pm$ 0.001 
			&5.08 $\pm$ 0.70 & (662/559=1.18)\\
			
			\hline \hline
		\end{tabular}
		\label{tab:Enspec_kerrbb_grs}
	}
\end{table*}


In this section, we present the results of broadband spectral modelling (Orbits 3826, 3830, 3841, 3860)
to constrain the mass and spin of the source. While doing this, we use the \texttt{kerrbb} model 
\citep{Li2005} along with the \texttt{simpl} model \citep{Steiner2009} of \texttt{XSPEC} \citep{Arnaud1996} to fit the spectra. The \texttt{kerrbb} model represents a thin, general relativistic multi temperature blackbody disc around a rotating black hole
whereas the \texttt{simpl} is a Comptonization model in which a fraction of the seed photons is scattered into a power-law distribution.
Following the spectral analysis method discussed in the previous section, we model the broadband 
observations with a combination of models as 
\texttt{Tbabs(simpl * kerbb $\times$ smedge)constant} that results into acceptable fit with $\chi^2_{\rm red} \sim 1$ (see Table \ref{tab:Enspec_kerrbb_grs}). The model fitted energy spectra for Orbits 3841 and 3860 are shown in Figure~\ref{fig:kerrbbspec}.
We find that the model fitted parameters of black hole mass $M_{\rm BH} =12.92 \pm 0.17~ M_{\odot}$, accretion rate ${\dot M} \sim 7.89 \times 10^{18}$~g/s and spin $a_{\rm k}=0.991 \pm 0.001$ for the observation of Orbit 3841. 
Similarly, we obtain $M_{\rm BH} =12.85 \pm 0.24 ~ M_{\odot}$, 
${\dot M} \sim 7.05 \times 10^{18}$ g/s and $a_{\rm k} = 0.995 \pm 0.001$  
for the observation of Orbit 3860. Interestingly, the accretion rate corresponding to non-detection of HFQPO (Orbit 3841) is higher than that in observations where HFQPO is detected. 
It is to be noted that two additional \texttt{gaussians} are used at $1.8$ keV and $2.2$ keV to account for the instrumental peaks at these energies, instead of using \texttt{gain fit}.

\begin{figure*}
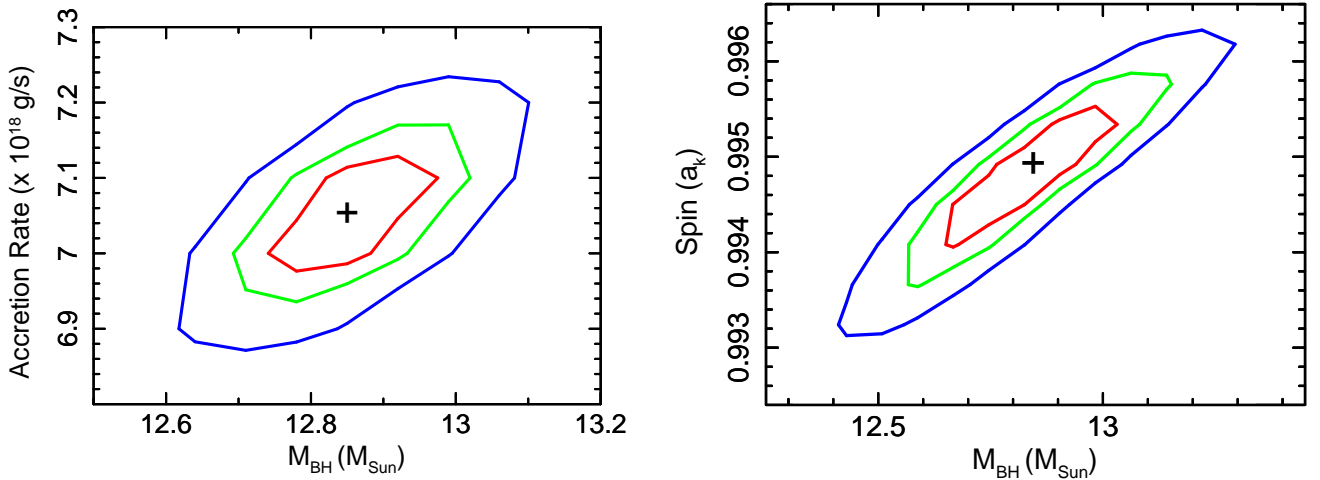

	\includegraphics[height=0.49\textwidth,angle=-90]{fig8a.eps}
	\includegraphics[height=0.49\textwidth,angle=-90]{fig8b.eps}
	\caption{The confidence contours of two parameters namely accretion rate (${\dot M}$) and mass of the black hole ($M_{\rm BH}$) for the observation during Orbit 3860 is shown on the left panel. On the right panel, we show the confidence contours for the mass ($M_{\rm BH}$) and spin ($a_{\rm k}$). The red, green and blue contours show delta fit statistic of 2.30 (68\%), 4.61 (90\%) and 9.21 (99\%), respectively.}
	\label{fig:confidence}
\end{figure*}

In Table \ref{tab:Enspec_kerrbb_grs}, we present the spectral fit parameters for all broadband observations, where column $1-8$ represent observation date with 
Orbit number, photon index ($\Gamma$), scattered fraction ($F_{\rm sca}$), accretion rate 
(${\dot M}$), black hole mass ($M_{\rm BH}$) in solar mass unit ($M_\odot$), 
spin ($a_{\rm k}$), \texttt{smedge} energy (${\rm E}_{\rm smedge}$) and $\chi^2_{\rm red}$ ($\chi^2/{\rm dof}$), respectively. The index ($\Gamma$) varies from $3.08 - 4.80$, whereas scattered fraction of seed photons is minimum ($F_{\rm sca} \sim 0.05$) when the HFQPO is absent and maximum ($F_{\rm sca} \sim 0.55$) when the HFQPO signal is most significant. 
Adopting the source mass as obtained from the spectral fitting (see Table \ref{tab:Enspec_kerrbb_grs}), and considering the source distance as $\sim 8.6$ kpc \citep{Reid2014}, we find that GRS 1915$+$105 accretes at supper-Eddington rate of $1.17-1.31~\dot{M}_{\rm Edd}$ during the GT phase {\it AstroSat} observations under consideration.

Further, we examine the dependence of accretion rate and spin on the mass of the source. Figure \ref{fig:confidence} (left) shows the confidence contours obtained for mass of the black hole and the accretion rate for the Orbit 3860. Similarly, in Figure \ref{fig:confidence} (right), we depict the confidence contours of mass and spin of the source for the same observation. It indicates that the black hole mass ($M_{\rm BH}$) lies in the range $12.85 \pm 0.24~M_\odot$ and spin ($a_{\rm k}$) lies in the range $0.993-0.996$. The red, green and blue curves represent contours of 68\%, 90\% and 99\% confidence, respectively.

Since our goal is to estimate black hole parameters, we have also attempted with the \texttt{kerrd} model \citep{Ebisawa2003}. The spectral fits yields a mass range of 11.66 - 12.69~${\rm M_{\odot}}$ and accretion rate of 7.05 to 7.75 $(\times 10^{18}~g/s)$. The mass estimate from \texttt{kerrd} model is marginally lower than the estimates from \texttt{kerrbb}, while the accretion rate estimate from both models are consistent. It may be noted that \texttt{kerrd} model assumes fixed spin value as 0.998, whereas the spin is treated as free parameter in \texttt{kerrbb} model, besides accretion rate and mass. Because of that we prefer \texttt{kerrbb} model over \texttt{kerrd} model and quote the parameters from \texttt{kerrbb} model as our final result.

\section{Discussion and Conclusion}

\begin{figure*}
	\begin{center}
		\includegraphics[width=\textwidth]{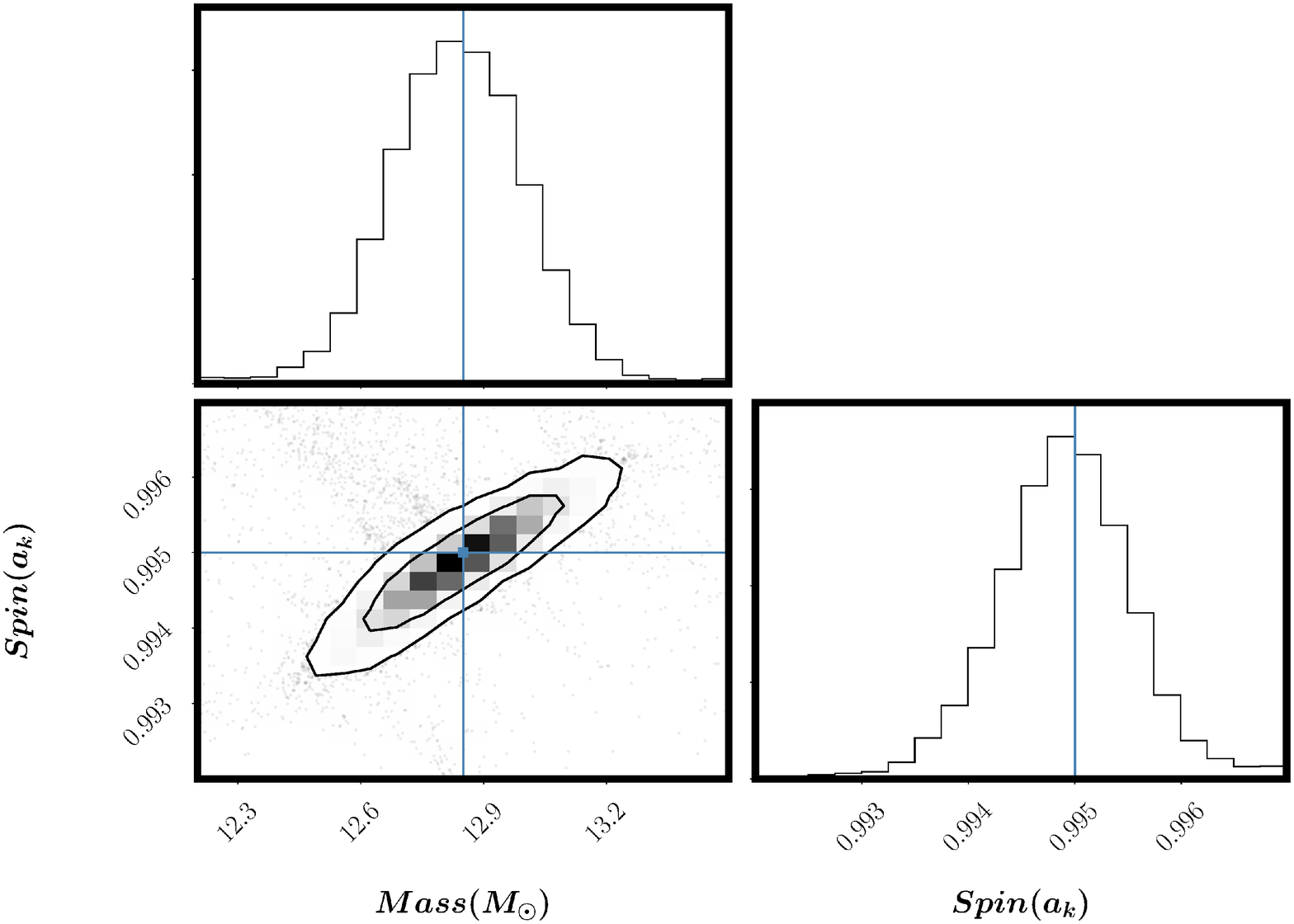} 
	\end{center}
	\caption{Contour plot of mass versus spin generated using the MCMC Hammer algorithm to determine errors on spin and mass of the black hole using data of Orbit 3860. The top panel shows mass distribution and the corner panel shows the contours
			of mass and spin parameters at 68\% and 90\% confidence. The right panel shows the spin distribution.}
	\label{fig:EMCEE}
\end{figure*}

\label{s:disc_grs}

In this paper, we study the spectro-temporal properties of GRS 1915$+$105 using GT phase observations of the {\it Astrosat} data. The CCDs indicate that the source exhibits $\delta$ class variability \citep{Belloni-etal00} with small hardness ratios (HR$_1 < 1.1,$ HR$_2 < 0.12$). During our observational campaign, GRS 1915$+$105 was in the soft state and significant variations in the count rate (7728 - 9865 cts/s), HR1 (0.75 - 0.92) and HR2 (0.04 - 0.08) are observed. We also find that HFQPOs disappear at high count rates and the disk fraction varies from 51\% to 69.7\%. This result indicates that the source presumably was in a time varying soft state during our campaign. 

We detect HFQPOs at $\sim 69$ Hz from the source during {\it AstroSat} observations. It is observed that the strength of the HFQPOs decreases and eventually disappears as the source count rate increases (see Figure \ref{fig:allqpos}) and hardness ratios decreases (see Figure \ref{fig:LCnCCD} and Table \ref{tab:laxpc_grs_qpo}). However, we have not observed significant change in the QPO frequency as the count rate varies. The dependence of QPO parameters on source intensity has been studied earlier by \cite{VanderKlis1985}, where they found a significant increase in frequency and decrease in strength of QPOs with increasing count rates. \cite{Cui2000} found that HFQPOs disappear with an increase in accretion rate in LMXB systems, which is also seen in our analysis. On the other hand, the disappearance of LFQPOs in LMXBs is shown to be associated with radio flares \citep{Fender2009, Nandi2013, Radhika-Nandi2014} and is followed by the subsequent softening of the energy spectra \citep{Rad2016}. During our observations, the hardness ratios of the source (see Table \ref{tab:laxpc_grs_qpo}) also indicate that the source becomes relatively softer when the HFQPOs are absent. 
From the energy dependent power spectra, it is evident that the HFQPOs are present only in the $6 - 25$~keV (see Figure \ref{fig:Endep3819}) but not in the $3 - 6$~keV and $25 - 60$~keV energy bands. Following \cite{Belloni2001}, we calculate the upper limit of the QPO significance in $3 - 6$ keV as $1.69$ and in $25 - 60$~keV as $0.44$. The rms amplitude of the HFQPOs in our observations is found to vary within $0.83 - 1.90$~\% and the frequencies of these HFQPOs lie in the range of $67.96-70.62$ Hz. \cite{Belloni-Altamirano13} also reported  HFQPO of frequency $67.4 \pm 2.0$ Hz, using \texttt{RXTE} data of the same source. Recently, \cite{Belloni2019} reported a variable HFQPO in the range between $67.4 - 72.3$~Hz, where higher phase lags are observed in harder regions. 

HFQPOs are detected in other BH-XRB sources, such as  GRO~J1655$-$40, XTE~J1550$-$564 and H~1743$-$322 \citep{Remillard-McClintock06}. 
Phase-lag studies of the 67~Hz QPO in GRS 1915$+$105 \citep{Cui1999,Mendez2013} indicates that the hard photons lag the soft photons. This is consistent with our scenario where soft radiation from the accretion disc is inverse-Comptonized within a compact corona around the black hole. A 66 Hz HFQPO was detected in the black hole binary source IGR J17091$-$3624 by \cite{Altamirano2012}. They have discussed the possibility that this HFQPO ($\sim 67$~Hz) detected in both GRS~1915$+$105 and IGR~J17091$-$3624 is linked to the physical processes that produce low-frequency structured variability exhibited by these two sources.

Several models have been proposed to explain the origin of these QPOs.
The relativistic precession model by \cite{Stella-Vietri98} considers that HFQPOs are associated to the orbital frequency as well as the nodal and the periastron precession frequencies. \cite{Rezolla2003} proposed a model where HFQPOs are attributed to the pressure mode oscillations of an accretion torus orbiting the black hole. Recently, in a theoretical attempt, \cite{Dihingia2019} showed that shock induced accretion solutions around a rapidly rotating stellar mass black hole are viable to account for such an HFQPO phenomenon. Interestingly, \cite{Varniere2020} explored the possibility of HFQPO generation resulting from vortices formed due to Rossby wave instability (RWI) at the inner edge of the accretion disc. As the centroid frequencies of the HFQPOs do not vary beyond a few percent, we conjecture that it perhaps originated from the vicinity of the source and therefore carries the imprint of the effect of strong gravity, namely the mass and spin of the black hole.

In order to explain the radiative properties of the accretion flow around the source, we carry out the broadband energy spectral modelling using combined {\it SXT} and {\it LAXPC} data in the $0.7 - 50$~keV energy range. Modelling of spectra with a multi-temperature disc blackbody (\texttt{diskbb}) along with a \texttt{powerlaw} indicates high disc contribution ($51 - 69.7\%$) to the total emitted radiative flux. We find relatively steeper photon index ($\sim 3$) and high disc temperature ($\sim 2.5$ keV) that eventually indicate the source belongs to the soft state.

Next, to examine the physical processes and the spectral parameters, we model the spectra using \texttt{nthComp} and \texttt{powerlaw}. The photon index ($\Gamma_{\rm nth}$) is seen to be small and remain nearly constant ($\Gamma_{\rm nth} \sim 1.7$) during the HFQPO observations whereas $\Gamma_{\rm nth}$ becomes large ($\Gamma _{\rm nth} \sim 2.45$) when HFQPO is not detected (see Table \ref{tab:bbspec_pheno}). It may be noted that the seed photon temperature does not change appreciably ($kT_{\rm bb} \sim 0.1-0.3$ keV) in these observations. We find electron temperature $kT_{\rm e}$ in the range $2.07 - 2.43$ keV ($i.e.$, an exponential roll over energy $\sim 7.3$ keV) and $1.73 < \Gamma_{\rm nth} < 2.45$ for broadband spectra. In addition, an extended corona seems to be present as well, because an additional \texttt{powerlaw} component with higher photon index ($\Gamma_{\rm PL} \sim 3$) is required to model the higher energy part of the spectrum (above $25$ keV). Indeed, it may be plausible that the accretion flow harbours a cool and compact central corona that not only exhibits HFQPOs, but also emits high energy photons till $\sim 25$ keV (see Figures \ref{fig:Endep3819} and \ref{fig:spec_bb_pheno}). Also, surrounding the inner compact corona, a diffuse but relatively hot extended `Compton' corona may be present which is likely to produce hard X-ray photons up to $\sim 50$ keV. The existence of such an extended coronal structure around GRS 1915$+$105 would be possible as the source may not swallow all the matter accreted at super-Eddington rate \citep{Done-etal2004,Punsly-etal2013}.

The broadband spectra of observations with relatively higher rms of HFQPOs indicate a weaker extended corona ($15.5<{\rm norm}_{\rm PL}<25.5$) in comparison to those spectral observations where HFQPOs are absent (${\rm norm}_{\rm PL}\sim 30$). Besides this, we notice from Table \ref{tab:bbspec_pheno} that the observations with significant HFQPOs have a stronger (${\rm norm}_{\rm nth} \sim 8.0$) and harder (${\rm \Gamma}_{\rm nth} \sim 1.73-1.79$) `compact corona' than the observation (Orbit 3841) without an HFQPO. This is an indication that the origin of the HFQPO is due to the oscillations of the `compact corona' that is represented by the \texttt{nthComp} model. The fact that the \texttt{powerlaw} is strong (${\rm norm}_{\rm PL} \sim 34.9$) and \texttt{nthComp} is comparatively weak (${\rm norm}_{\rm nth} \sim 3.9$) for Orbit 3839 which has the HFQPO with lowest rms ($0.83$\%) also supports our argument that the HFQPO is the result of a compact oscillating corona represented by the \texttt{nthComp}.
\cite{Dihingia2019} indicated that HFQPOs at $\sim 67$ Hz in GRS 1915$+$105 possibly originated due to the quasi-periodic modulation of a very compact inner region of the disc. It is noteworthy that the energy spectra of the source GRO~J1655$-$40 corresponding to the HFQPOs of frequency 300~Hz \citep{Remillard1999} and 450~Hz \citep{Strohmayer2001b} possess strong Comptonized contribution extending beyond $100$ keV \citep{Aktar2018}.
%
%
Meanwhile, \cite{Remillard2002} also pointed out that the HFQPOs perhaps originated because of the modulation of the compact Compton corona. Overall, we conjecture that the observed HFQPOs are the manifestation of the oscillations of a `hot' and compact post-shock corona \citep{Aktar2018,Dihingia2019}.

In addition, we model the broadband energy spectra of all the observations using \texttt{kerrbb} model and estimate mass and spin of the source along with the flow accretion rate. Our findings reveal that the mass of the source ($M_{\rm BH}$) lies in the range $12.44-13.09 ~M_{\odot}$ (see Table \ref{tab:Enspec_kerrbb_grs}). This estimate is consistent and better constrained compared to the earlier estimates \citep{Greiner-etal01,Reid2014}. We find the spin of the source ($a_{\rm k}$) in the range $0.990-0.997$ with $90\%$ confidence whereas the previous estimate of spin value was reported as $a_{\rm k}=0.98 \pm 0.01$ \citep{Rebecca2006,Blum-etal09,Miller-etal13}. The fraction of Compton up-scattering (${\rm F}_{\rm sca}$) is found to be as high as $\sim 0.55$ when a strong HFQPO is detected (Orbit 3860), whereas ${\rm F}_{\rm sca}$ is low ($0.05$) during the absence of HFQPO (Orbit 3841). The index of the \texttt{simpl} convolution model is the lowest ($\sim 3.09$) during the non-detection of HFQPO and highest ($4.80$) during the detection of a strong HFQPO (Orbit 3860). Our results also indicate that the source accretes at super-Eddington rate in the range $1.17-1.31 ~ {\dot M}_{\rm Edd}$ during GT phase of {\it AstroSat} observations.

In order to improve the error estimation, we have carried out Markov Chain Monte Carlo (MCMC) simulations. We use the Goodman-Weare chain algorithm \citep{GoodmanWeare2010} in \texttt{XSPEC} \citep{Arnaud1996} with \texttt{walkers} parameter set to 32. The chain length is taken as 15,000 and burn length is fixed to 5,000. However, there is no significant improvement in the error estimations. We also cross-check our error estimations using the MCMC hammer algorithm (emcee\footnote{https://emcee.readthedocs.io/en/stable/\#}) based on \cite{Foreman2013}. These results are limited by the unavailability of an analytical approximation for \texttt{kerrbb} model. 
The results from emcee estimation of spin and mass parameters for Orbit 3860 are shown in Figure \ref{fig:EMCEE}. The estimate of mass of the black hole for this observation is $12.47 - 13.23~M_{\odot}$ and the spin is estimated to be $0.993- 0.996$ at 90\% confidence level. It is to be noted that the errors quoted in this paper are purely statistical and systematics are not accounted for. The top and right panels show mass and spin distribution respectively, whereas the corner panel shows the confidence contours considering both mass and spin. The contour plot from MCMC simulations and the contour generated from \texttt{XSPEC} (see right panel of Figure \ref{fig:confidence}) using \texttt{steppar} functions are consistent.

Finally, we emphasize that GRS 1915$+$105 is a maximally rotating, comparatively higher mass X-ray binary source accreting at supper-Eddington rate and exhibiting HFQPO features. In addition, we point out that for the first time to the best of our knowledge, the mass, spin, accretion rate and HFQPOs of GRS 1915$+$105 are concurrently examined and reported using {\it AstroSat} observations.


\section*{Acknowledgments}
We thank the anonymous reviewer for his/her suggestions and comments that helped us to improve the quality of this manuscript.
AN, VKA, MCR thank DD, PDMSA and Director, URSC for encouragement and continuous support to 
carry out this research. SH thanks Department of Physics, IIT Guwahati for providing the facilities to complete part of this work. This research made use of the data obtained through GT phase of 
{\it AstroSat} observations. The authors thank the SXT-POC of TIFR and the
{\it LAXPC} team of IUCAA and TIFR for providing the data extraction software for the respective instruments.

\section*{Data Availability}
Data used for this publication are currently available at the Astrobrowse (AstroSat archive) website (\url{https://astrobrowse.issdc.gov.in/astro\_archive/archive}) of the Indian Space Science Data Center (ISSDC).

\bibliography{refs}

\begin{thebibliography}{}
\makeatletter
\relax
\def\mn@urlcharsother{\let\do\@makeother \do\$\do\&\do\#\do\^\do\_\do\%\do\~}
\def\mn@doi{\begingroup\mn@urlcharsother \@ifnextchar [ {\mn@doi@}
  {\mn@doi@[]}}
\def\mn@doi@[#1]#2{\def\@tempa{#1}\ifx\@tempa\@empty \href
  {http://dx.doi.org/#2} {doi:#2}\else \href {http://dx.doi.org/#2} {#1}\fi
  \endgroup}
\def\mn@eprint#1#2{\mn@eprint@#1:#2::\@nil}
\def\mn@eprint@arXiv#1{\href {http://arxiv.org/abs/#1} {{\tt arXiv:#1}}}
\def\mn@eprint@dblp#1{\href {http://dblp.uni-trier.de/rec/bibtex/#1.xml}
  {dblp:#1}}
\def\mn@eprint@#1:#2:#3:#4\@nil{\def\@tempa {#1}\def\@tempb {#2}\def\@tempc
  {#3}\ifx \@tempc \@empty \let \@tempc \@tempb \let \@tempb \@tempa \fi \ifx
  \@tempb \@empty \def\@tempb {arXiv}\fi \@ifundefined
  {mn@eprint@\@tempb}{\@tempb:\@tempc}{\expandafter \expandafter \csname
  mn@eprint@\@tempb\endcsname \expandafter{\@tempc}}}

\bibitem[\protect\citeauthoryear{{Agrawal} et~al.,}{{Agrawal}
  et~al.}{2017}]{Agarwal-etal17}
{Agrawal} P.~C.,  et~al., 2017, \mn@doi [Journal of Astrophysics and Astronomy]
  {10.1007/s12036-017-9451-z}, \href
  {http://adsabs.harvard.edu/abs/2017JApA...38...30A} {38, 30}

\bibitem[\protect\citeauthoryear{{Agrawal}, {Nandi}, {Girish}  \&
  {Ramadevi}}{{Agrawal} et~al.}{2018}]{Agrawal-etal2018}
{Agrawal} V.~K.,  {Nandi} A.,  {Girish} V.,   {Ramadevi} M.~C.,  2018, \mn@doi
  [\mnras] {10.1093/mnras/sty1005}, \href
  {http://adsabs.harvard.edu/abs/2018MNRAS.477.5437A} {477, 5437}

\bibitem[\protect\citeauthoryear{{Aktar}, {Das}, {Nandi}  \&
  {Sreehari}}{{Aktar} et~al.}{2018}]{Aktar2018}
{Aktar} R.,  {Das} S.,  {Nandi} A.,   {Sreehari} H.,  2018, \mn@doi [Journal of
  Astrophysics and Astronomy] {10.1007/s12036-017-9507-0}, \href
  {https://ui.adsabs.harvard.edu/abs/2018JApA...39...17A} {39, 17}

\bibitem[\protect\citeauthoryear{{Alam}, {Dewangan}, {Belloni}, {Mukherjee}  \&
  {Jhingan}}{{Alam} et~al.}{2014}]{Alam2014}
{Alam} M.~S.,  {Dewangan} G.~C.,  {Belloni} T.,  {Mukherjee} D.,   {Jhingan}
  S.,  2014, \mn@doi [\mnras] {10.1093/mnras/stu2048}, \href
  {https://ui.adsabs.harvard.edu/abs/2014MNRAS.445.4259A} {445, 4259}

\bibitem[\protect\citeauthoryear{{Altamirano} \& {Belloni}}{{Altamirano} \&
  {Belloni}}{2012}]{Altamirano2012}
{Altamirano} D.,  {Belloni} T.,  2012, \mn@doi [\apjl]
  {10.1088/2041-8205/747/1/L4}, \href
  {https://ui.adsabs.harvard.edu/abs/2012ApJ...747L...4A} {747, L4}

\bibitem[\protect\citeauthoryear{{Antia} et~al.,}{{Antia}
  et~al.}{2017}]{Antia-etal17}
{Antia} H.~M.,  et~al., 2017, \mn@doi [\apjs] {10.3847/1538-4365/aa7a0e}, \href
  {http://adsabs.harvard.edu/abs/2017ApJS..231...10A} {231, 10}

\bibitem[\protect\citeauthoryear{{Arnaud}}{{Arnaud}}{1996}]{Arnaud1996}
{Arnaud} K.~A.,  1996, {XSPEC: The First Ten Years}.
p.~17

\bibitem[\protect\citeauthoryear{{Bachetti} et~al.,}{{Bachetti}
  et~al.}{2015}]{Bachetti2015}
{Bachetti} M.,  et~al., 2015, \mn@doi [\apj] {10.1088/0004-637X/800/2/109},
  \href {http://adsabs.harvard.edu/abs/2015ApJ...800..109B} {800, 109}

\bibitem[\protect\citeauthoryear{{Belloni} \& {Altamirano}}{{Belloni} \&
  {Altamirano}}{2013}]{Belloni-Altamirano13}
{Belloni} T.~M.,  {Altamirano} D.,  2013, \mn@doi [\mnras]
  {10.1093/mnras/stt500}, \href
  {http://adsabs.harvard.edu/abs/2013MNRAS.432...10B} {432, 10}

\bibitem[\protect\citeauthoryear{{Belloni} \& {Motta}}{{Belloni} \&
  {Motta}}{2016}]{Belloni-Motta16}
{Belloni} T.~M.,  {Motta} S.~E.,  2016, in {Bambi} C.,  ed.,  Astrophysics and
  Space Science Library Vol. 440, Astrophysics of Black Holes: From Fundamental
  Aspects to Latest Developments. p.~61 (\mn@eprint {arXiv} {1603.07872}),
  \mn@doi{10.1007/978-3-662-52859-4_2}

\bibitem[\protect\citeauthoryear{{Belloni}, {Klein-Wolt}, {M{\'e}ndez}, {van
  der Klis}  \& {van Paradijs}}{{Belloni} et~al.}{2000}]{Belloni-etal00}
{Belloni} T.,  {Klein-Wolt} M.,  {M{\'e}ndez} M.,  {van der Klis} M.,   {van
  Paradijs} J.,  2000, \aap, \href
  {http://adsabs.harvard.edu/abs/2000A%26A...355..271B} {355, 271}

\bibitem[\protect\citeauthoryear{{Belloni}, {M{\'e}ndez}  \&
  {S{\'a}nchez-Fern{\'a}ndez}}{{Belloni} et~al.}{2001}]{Belloni2001}
{Belloni} T.,  {M{\'e}ndez} M.,   {S{\'a}nchez-Fern{\'a}ndez} C.,  2001,
  \mn@doi [\aap] {10.1051/0004-6361:20010480}, \href
  {https://ui.adsabs.harvard.edu/abs/2001A&A...372..551B} {372, 551}

\bibitem[\protect\citeauthoryear{{Belloni}, {Homan}, {Casella}, {van der Klis},
  {Nespoli}, {Lewin}, {Miller}  \& {M{\'e}ndez}}{{Belloni}
  et~al.}{2005}]{Belloni2005}
{Belloni} T.,  {Homan} J.,  {Casella} P.,  {van der Klis} M.,  {Nespoli} E.,
  {Lewin} W.~H.~G.,  {Miller} J.~M.,   {M{\'e}ndez} M.,  2005, \mn@doi [\aap]
  {10.1051/0004-6361:20042457}, \href
  {https://ui.adsabs.harvard.edu/abs/2005A&A...440..207B} {440, 207}

\bibitem[\protect\citeauthoryear{{Belloni}, {Soleri}, {Casella}, {M{\'e}ndez}
  \& {Migliari}}{{Belloni} et~al.}{2006}]{Belloni2006}
{Belloni} T.,  {Soleri} P.,  {Casella} P.,  {M{\'e}ndez} M.,   {Migliari} S.,
  2006, \mn@doi [\mnras] {10.1111/j.1365-2966.2006.10286.x}, \href
  {https://ui.adsabs.harvard.edu/abs/2006MNRAS.369..305B} {369, 305}

\bibitem[\protect\citeauthoryear{Belloni, Bhattacharya, Caccese, Bhalerao,
  Vadawale  \& Yadav}{Belloni et~al.}{2019}]{Belloni2019}
Belloni T.~M.,  Bhattacharya D.,  Caccese P.,  Bhalerao V.,  Vadawale S.,
  Yadav J.~S.,  2019, \mn@doi [\mnras] {10.1093/mnras/stz2143}, 489, 1037

\bibitem[\protect\citeauthoryear{{Blum}, {Miller}, {Fabian}, {Miller}, {Homan},
  {van der Klis}, {Cackett}  \& {Reis}}{{Blum} et~al.}{2009}]{Blum-etal09}
{Blum} J.~L.,  {Miller} J.~M.,  {Fabian} A.~C.,  {Miller} M.~C.,  {Homan} J.,
  {van der Klis} M.,  {Cackett} E.~M.,   {Reis} R.~C.,  2009, \mn@doi [\apj]
  {10.1088/0004-637X/706/1/60}, \href
  {http://adsabs.harvard.edu/abs/2009ApJ...706...60B} {706, 60}

\bibitem[\protect\citeauthoryear{{Castro-Tirado}, {Brandt}  \&
  {Lund}}{{Castro-Tirado} et~al.}{1992}]{Castro1992}
{Castro-Tirado} A.~J.,  {Brandt} S.,   {Lund} N.,  1992, IAU, \href
  {http://adsabs.harvard.edu/abs/1992IAUC.5590....2C} {5590}

\bibitem[\protect\citeauthoryear{{Chakrabarti} \& {Titarchuk}}{{Chakrabarti} \&
  {Titarchuk}}{1995}]{CT1995}
{Chakrabarti} S.,  {Titarchuk} L.~G.,  1995, \mn@doi [\apj] {10.1086/176610},
  \href {https://ui.adsabs.harvard.edu/abs/1995ApJ...455..623C} {455, 623}

\bibitem[\protect\citeauthoryear{{Cui}}{{Cui}}{1999}]{Cui1999}
{Cui} W.,  1999, \mn@doi [\apjl] {10.1086/312296}, \href
  {https://ui.adsabs.harvard.edu/abs/1999ApJ...524L..59C} {524, L59}

\bibitem[\protect\citeauthoryear{{Cui}}{{Cui}}{2000}]{Cui2000}
{Cui} W.,  2000, \mn@doi [\apjl] {10.1086/312646}, \href
  {https://ui.adsabs.harvard.edu/abs/2000ApJ...534L..31C} {534, L31}

\bibitem[\protect\citeauthoryear{{Cui}, {Zhang}  \& {Chen}}{{Cui}
  et~al.}{1998}]{Cui-etal98}
{Cui} W.,  {Zhang} S.~N.,   {Chen} W.,  1998, \mn@doi [\apjl] {10.1086/311092},
  \href {http://adsabs.harvard.edu/abs/1998ApJ...492L..53C} {492, L53}

\bibitem[\protect\citeauthoryear{{Dihingia}, {Das}, {Maity}  \& {Nand
  i}}{{Dihingia} et~al.}{2019}]{Dihingia2019}
{Dihingia} I.~K.,  {Das} S.,  {Maity} D.,   {Nand i} A.,  2019, \mn@doi
  [\mnras] {10.1093/mnras/stz1933}, \href
  {https://ui.adsabs.harvard.edu/abs/2019MNRAS.488.2412D} {488, 2412}

\bibitem[\protect\citeauthoryear{{Done}, {Wardzi{\'n}ski}  \&
  {Gierli{\'n}ski}}{{Done} et~al.}{2004}]{Done-etal2004}
{Done} C.,  {Wardzi{\'n}ski} G.,   {Gierli{\'n}ski} M.,  2004, \mn@doi [\mnras]
  {10.1111/j.1365-2966.2004.07545.x}, \href
  {http://adsabs.harvard.edu/abs/2004MNRAS.349..393D} {349, 393}

\bibitem[\protect\citeauthoryear{{Ebisawa}, {{\.Z}ycki}, {Kubota}, {Mizuno}  \&
  {Watarai}}{{Ebisawa} et~al.}{2003}]{Ebisawa2003}
{Ebisawa} K.,  {{\.Z}ycki} P.,  {Kubota} A.,  {Mizuno} T.,   {Watarai} K.-y.,
  2003, \mn@doi [\apj] {10.1086/378586}, \href
  {https://ui.adsabs.harvard.edu/abs/2003ApJ...597..780E} {597, 780}

\bibitem[\protect\citeauthoryear{{Fender}, {Homan}  \& {Belloni}}{{Fender}
  et~al.}{2009}]{Fender2009}
{Fender} R.~P.,  {Homan} J.,   {Belloni} T.~M.,  2009, \mn@doi [\mnras]
  {10.1111/j.1365-2966.2009.14841.x}, \href
  {https://ui.adsabs.harvard.edu/abs/2009MNRAS.396.1370F} {396, 1370}

\bibitem[\protect\citeauthoryear{{Feroci}, {Matt}, {Pooley}, {Costa}, {Tavani}
  \& {Belloni}}{{Feroci} et~al.}{1999}]{Feroci1999}
{Feroci} M.,  {Matt} G.,  {Pooley} G.,  {Costa} E.,  {Tavani} M.,   {Belloni}
  T.,  1999, \aap, \href
  {https://ui.adsabs.harvard.edu/abs/1999A&A...351..985F} {351, 985}

\bibitem[\protect\citeauthoryear{{Foreman-Mackey}, {Hogg}, {Lang}  \&
  {Goodman}}{{Foreman-Mackey} et~al.}{2013}]{Foreman2013}
{Foreman-Mackey} D.,  {Hogg} D.~W.,  {Lang} D.,   {Goodman} J.,  2013, \mn@doi
  [Publications of the Astronomical Society of the Pacific] {10.1086/670067},
  \href {https://ui.adsabs.harvard.edu/abs/2013PASP..125..306F} {125, 306}

\bibitem[\protect\citeauthoryear{{Garc{\'{\i}}a} et~al.,}{{Garc{\'{\i}}a}
  et~al.}{2014}]{Garcia2014}
{Garc{\'{\i}}a} J.,  et~al., 2014, \mn@doi [\apj] {10.1088/0004-637X/782/2/76},
  \href {http://adsabs.harvard.edu/abs/2014ApJ...782...76G} {782, 76}

\bibitem[\protect\citeauthoryear{{Goodman} \& {Weare}}{{Goodman} \&
  {Weare}}{2010}]{GoodmanWeare2010}
{Goodman} J.,  {Weare} J.,  2010, \mn@doi [Communications in Applied
  Mathematics and Computational Science] {10.2140/camcos.2010.5.65}, \href
  {https://ui.adsabs.harvard.edu/abs/2010CAMCS...5...65G} {5, 65}

\bibitem[\protect\citeauthoryear{{Greiner}, {Cuby}  \& {McCaughrean}}{{Greiner}
  et~al.}{2001}]{Greiner-etal01}
{Greiner} J.,  {Cuby} J.~G.,   {McCaughrean} M.~J.,  2001, \mn@doi [\nat]
  {10.1038/35107019}, \href {http://adsabs.harvard.edu/abs/2001Natur.414..522G}
  {414, 522}

\bibitem[\protect\citeauthoryear{{Grove}, {Johnson}, {Kroeger},
  {McNaron-Brown}, {Skibo}  \& {Phlips}}{{Grove} et~al.}{1998}]{Grove1998}
{Grove} J.~E.,  {Johnson} W.~N.,  {Kroeger} R.~A.,  {McNaron-Brown} K.,
  {Skibo} J.~G.,   {Phlips} B.~F.,  1998, \mn@doi [\apj] {10.1086/305746},
  \href {https://ui.adsabs.harvard.edu/abs/1998ApJ...500..899G} {500, 899}

\bibitem[\protect\citeauthoryear{{Hannikainen} et~al.,}{{Hannikainen}
  et~al.}{2005}]{Hann2005}
{Hannikainen} D.~C.,  et~al., 2005, \mn@doi [\aap]
  {10.1051/0004-6361:20042250}, \href
  {https://ui.adsabs.harvard.edu/abs/2005A&A...435..995H} {435, 995}

\bibitem[\protect\citeauthoryear{{Iyer}, {Nandi}  \& {Mandal}}{{Iyer}
  et~al.}{2015}]{Iyer2015}
{Iyer} N.,  {Nandi} A.,   {Mandal} S.,  2015, \mn@doi [\apj]
  {10.1088/0004-637X/807/1/108}, \href
  {https://ui.adsabs.harvard.edu/abs/2015ApJ...807..108I} {807, 108}

\bibitem[\protect\citeauthoryear{{Klein-Wolt}, {Fender}, {Pooley}, {Belloni},
  {Migliari}, {Morgan}  \& {van der Klis}}{{Klein-Wolt}
  et~al.}{2002}]{Klein_wolt_etal2002}
{Klein-Wolt} M.,  {Fender} R.~P.,  {Pooley} G.~G.,  {Belloni} T.,  {Migliari}
  S.,  {Morgan} E.~H.,   {van der Klis} M.,  2002, \mn@doi [\mnras]
  {10.1046/j.1365-8711.2002.05223.x}, \href
  {https://ui.adsabs.harvard.edu/abs/2002MNRAS.331..745K} {331, 745}

\bibitem[\protect\citeauthoryear{{Leahy} \& {Chen}}{{Leahy} \&
  {Chen}}{2019}]{Leahy2019}
{Leahy} D.~A.,  {Chen} Y.,  2019, \mn@doi [\apj] {10.3847/1538-4357/aaf8a9},
  \href {http://adsabs.harvard.edu/abs/2019ApJ...871..152L} {871, 152}

\bibitem[\protect\citeauthoryear{{Leahy}, {Darbro}, {Elsner}, {Weisskopf},
  {Sutherland}, {Kahn}  \& {Grindlay}}{{Leahy} et~al.}{1983}]{Leahy1983}
{Leahy} D.~A.,  {Darbro} W.,  {Elsner} R.~F.,  {Weisskopf} M.~C.,  {Sutherland}
  P.~G.,  {Kahn} S.,   {Grindlay} J.~E.,  1983, \mn@doi [\apj]
  {10.1086/160766}, \href {http://adsabs.harvard.edu/abs/1983ApJ...266..160L}
  {266, 160}

\bibitem[\protect\citeauthoryear{{Li}, {Zimmerman}, {Narayan}  \&
  {McClintock}}{{Li} et~al.}{2005}]{Li2005}
{Li} L.-X.,  {Zimmerman} E.~R.,  {Narayan} R.,   {McClintock} J.~E.,  2005,
  \mn@doi [\apjs] {10.1086/428089}, \href
  {https://ui.adsabs.harvard.edu/abs/2005ApJS..157..335L} {157, 335}

\bibitem[\protect\citeauthoryear{{M{\'e}ndez}, {Altamirano}, {Belloni}  \&
  {Sanna}}{{M{\'e}ndez} et~al.}{2013}]{Mendez2013}
{M{\'e}ndez} M.,  {Altamirano} D.,  {Belloni} T.,   {Sanna} A.,  2013, \mn@doi
  [\mnras] {10.1093/mnras/stt1431}, \href
  {https://ui.adsabs.harvard.edu/abs/2013MNRAS.435.2132M} {435, 2132}

\bibitem[\protect\citeauthoryear{{Merloni}, {Vietri}, {Stella}  \&
  {Bini}}{{Merloni} et~al.}{1999}]{Merloni-etal99}
{Merloni} A.,  {Vietri} M.,  {Stella} L.,   {Bini} D.,  1999, \mn@doi [\mnras]
  {10.1046/j.1365-8711.1999.02307.x}, \href
  {http://adsabs.harvard.edu/abs/1999MNRAS.304..155M} {304, 155}

\bibitem[\protect\citeauthoryear{{Miller} et~al.,}{{Miller}
  et~al.}{2013}]{Miller-etal13}
{Miller} J.~M.,  et~al., 2013, \mn@doi [\apjl] {10.1088/2041-8205/775/2/L45},
  \href {http://adsabs.harvard.edu/abs/2013ApJ...775L..45M} {775, L45}

\bibitem[\protect\citeauthoryear{{Morgan}, {Remillard}  \& {Greiner}}{{Morgan}
  et~al.}{1997}]{Morgan1997}
{Morgan} E.~H.,  {Remillard} R.~A.,   {Greiner} J.,  1997, \mn@doi [\apj]
  {10.1086/304191}, \href
  {https://ui.adsabs.harvard.edu/abs/1997ApJ...482..993M} {482, 993}

\bibitem[\protect\citeauthoryear{{Nandi}, {Debnath}, {Mandal}  \&
  {Chakrabarti}}{{Nandi} et~al.}{2012}]{Nandi-etal2012}
{Nandi} A.,  {Debnath} D.,  {Mandal} S.,   {Chakrabarti} S.~K.,  2012, \mn@doi
  [\aap] {10.1051/0004-6361/201117844}, \href
  {http://adsabs.harvard.edu/abs/2012A%26A...542A..56N} {542, A56}

\bibitem[\protect\citeauthoryear{{Nandi}, {Radhika}  \& {Seetha}}{{Nandi}
  et~al.}{2013}]{Nandi2013}
{Nandi} A.,  {Radhika} D.,   {Seetha} S.,  2013, in Astronomical Society of
  India Conference Series. pp 71--77 (\mn@eprint {arXiv} {1308.4567})

\bibitem[\protect\citeauthoryear{{Nowak}, {Wagoner}, {Begelman}  \&
  {Lehr}}{{Nowak} et~al.}{1997}]{Nowak-etal97}
{Nowak} M.~A.,  {Wagoner} R.~V.,  {Begelman} M.~C.,   {Lehr} D.~E.,  1997,
  \mn@doi [\apjl] {10.1086/310534}, \href
  {http://adsabs.harvard.edu/abs/1997ApJ...477L..91N} {477, L91}

\bibitem[\protect\citeauthoryear{{Punsly} \& {Rodriguez}}{{Punsly} \&
  {Rodriguez}}{2013}]{Punsly-etal2013}
{Punsly} B.,  {Rodriguez} J.,  2013, \mn@doi [\apj]
  {10.1088/0004-637X/764/2/173}, \href
  {https://ui.adsabs.harvard.edu/abs/2013ApJ...764..173P} {764, 173}

\bibitem[\protect\citeauthoryear{{Radhika} \& {Nandi}}{{Radhika} \&
  {Nandi}}{2014}]{Radhika-Nandi2014}
{Radhika} D.,  {Nandi} A.,  2014, \mn@doi [Advances in Space Research]
  {10.1016/j.asr.2014.06.039}, \href
  {https://ui.adsabs.harvard.edu/abs/2014AdSpR..54.1678R} {54, 1678}

\bibitem[\protect\citeauthoryear{{Radhika}, {Nandi}, {Agrawal}  \&
  {Seetha}}{{Radhika} et~al.}{2016}]{Rad2016}
{Radhika} D.,  {Nandi} A.,  {Agrawal} V.~K.,   {Seetha} S.,  2016, \mn@doi
  [\mnras] {10.1093/mnras/stw1239}, \href
  {https://ui.adsabs.harvard.edu/abs/2016MNRAS.460.4403R} {460, 4403}

\bibitem[\protect\citeauthoryear{{Radhika}, {Sreehari}, {Nandi}, {Iyer}  \&
  {Mand al}}{{Radhika} et~al.}{2018}]{Rad2018}
{Radhika} D.,  {Sreehari} H.,  {Nandi} A.,  {Iyer} N.,   {Mand al} S.,  2018,
  \mn@doi [\apss] {10.1007/s10509-018-3411-1}, \href
  {https://ui.adsabs.harvard.edu/abs/2018Ap&SS.363..189D} {363, 189}

\bibitem[\protect\citeauthoryear{Ratti, Belloni  \& Motta}{Ratti
  et~al.}{2012}]{Ratti2012}
Ratti E.~M.,  Belloni T.~M.,   Motta S.~E.,  2012, \mn@doi [Monthly Notices of
  the Royal Astronomical Society] {10.1111/j.1365-2966.2012.20906.x}, 423, 694

\bibitem[\protect\citeauthoryear{{Rebusco}}{{Rebusco}}{2008}]{Rebusco-08}
{Rebusco} P.,  2008, \mn@doi [\nar] {10.1016/j.newar.2008.03.015}, \href
  {http://adsabs.harvard.edu/abs/2008NewAR..51..855R} {51, 855}

\bibitem[\protect\citeauthoryear{{Reid}, {McClintock}, {Steiner}, {Steeghs},
  {Remillard}, {Dhawan}  \& {Narayan}}{{Reid} et~al.}{2014}]{Reid2014}
{Reid} M.~J.,  {McClintock} J.~E.,  {Steiner} J.~F.,  {Steeghs} D.,
  {Remillard} R.~A.,  {Dhawan} V.,   {Narayan} R.,  2014, \mn@doi [\apj]
  {10.1088/0004-637X/796/1/2}, \href
  {https://ui.adsabs.harvard.edu/abs/2014ApJ...796....2R} {796, 2}

\bibitem[\protect\citeauthoryear{{Remillard} \& {McClintock}}{{Remillard} \&
  {McClintock}}{2006}]{Remillard-McClintock06}
{Remillard} R.~A.,  {McClintock} J.~E.,  2006, \mn@doi [\araa]
  {10.1146/annurev.astro.44.051905.092532}, \href
  {http://adsabs.harvard.edu/abs/2006ARA%26A..44...49R} {44, 49}

\bibitem[\protect\citeauthoryear{{Remillard}, {Morgan}, {McClintock}, {Bailyn}
  \& {Orosz}}{{Remillard} et~al.}{1999}]{Remillard1999}
{Remillard} R.~A.,  {Morgan} E.~H.,  {McClintock} J.~E.,  {Bailyn} C.~D.,
  {Orosz} J.~A.,  1999, \mn@doi [\apj] {10.1086/307606}, \href
  {https://ui.adsabs.harvard.edu/abs/1999ApJ...522..397R} {522, 397}

\bibitem[\protect\citeauthoryear{{Remillard}, {Muno}, {McClintock}  \&
  {Orosz}}{{Remillard} et~al.}{2002}]{Remillard2002}
{Remillard} R.~A.,  {Muno} M.~P.,  {McClintock} J.~E.,   {Orosz} J.~A.,  2002,
  \mn@doi [\apj] {10.1086/343791}, \href
  {https://ui.adsabs.harvard.edu/abs/2002ApJ...580.1030R} {580, 1030}

\bibitem[\protect\citeauthoryear{{Rezzolla}, {Yoshida}, {Maccarone}  \&
  {Zanotti}}{{Rezzolla} et~al.}{2003}]{Rezolla2003}
{Rezzolla} L.,  {Yoshida} S.,  {Maccarone} T.~J.,   {Zanotti} O.,  2003,
  \mn@doi [\mnras] {10.1046/j.1365-8711.2003.07018.x}, \href
  {https://ui.adsabs.harvard.edu/abs/2003MNRAS.344L..37R} {344, L37}

\bibitem[\protect\citeauthoryear{{Seward} \& {Charles}}{{Seward} \&
  {Charles}}{2010}]{Seward2010}
{Seward} F.~D.,  {Charles} P.~A.,  2010, {Exploring the X-ray Universe}.
Cambridge University Press, 2010.~ISBN: 9780521884839

\bibitem[\protect\citeauthoryear{{Shafee}, {McClintock}, {Narayan},
  {Remillard}, {Davis}  \& {Li}}{{Shafee} et~al.}{2006}]{Rebecca2006}
{Shafee} R.,  {McClintock} J.~E.,  {Narayan} R.,  {Remillard} R.~A.,  {Davis}
  S.~W.,   {Li} L.,  2006, in AAS/High Energy Astrophysics Division \#9.
  AAS/High Energy Astrophysics Division.
p.~1.86

\bibitem[\protect\citeauthoryear{{Shakura} \& {Sunyaev}}{{Shakura} \&
  {Sunyaev}}{1973}]{ShaSu1973}
{Shakura} N.~I.,  {Sunyaev} R.~A.,  1973, \aap, \href
  {http://adsabs.harvard.edu/abs/1973A%26A....24..337S} {24, 337}

\bibitem[\protect\citeauthoryear{{Singh}, {Stewart}, {Westergaard},
  {Bhattacharayya}  \& {Chandra et al.}}{{Singh} et~al.}{2017}]{Singh2017}
{Singh} K.~P.,  {Stewart} G.~C.,  {Westergaard} N.~J.,  {Bhattacharayya} S.,
  {Chandra et al.} S.,  2017, \mn@doi [Journal of Astrophysics and Astronomy]
  {10.1007/s12036-017-9448-7}, \href
  {http://adsabs.harvard.edu/abs/2017JApA...38...29S} {38, 29}

\bibitem[\protect\citeauthoryear{{Sreehari}, {Ravishankar}, {Iyer}, {Agrawal},
  {Katoch}, {Mandal}  \& {Nand i}}{{Sreehari} et~al.}{2019}]{Sreehari-etal2019}
{Sreehari} H.,  {Ravishankar} B.~T.,  {Iyer} N.,  {Agrawal} V.~K.,  {Katoch}
  T.~B.,  {Mandal} S.,   {Nand i} A.,  2019, \mn@doi [\mnras]
  {10.1093/mnras/stz1327}, \href
  {https://ui.adsabs.harvard.edu/abs/2019MNRAS.487..928S} {487, 928}

\bibitem[\protect\citeauthoryear{{Stefanov}}{{Stefanov}}{2014}]{Stefanov14}
{Stefanov} I.~Z.,  2014, \mn@doi [\mnras] {10.1093/mnras/stu1602}, \href
  {http://adsabs.harvard.edu/abs/2014MNRAS.444.2178S} {444, 2178}

\bibitem[\protect\citeauthoryear{{Steiner}, {Narayan}, {McClintock}  \&
  {Ebisawa}}{{Steiner} et~al.}{2009}]{Steiner2009}
{Steiner} J.~F.,  {Narayan} R.,  {McClintock} J.~E.,   {Ebisawa} K.,  2009,
  \mn@doi [\pasp] {10.1086/648535}, \href
  {https://ui.adsabs.harvard.edu/abs/2009PASP..121.1279S} {121, 1279}

\bibitem[\protect\citeauthoryear{{Stella} \& {Vietri}}{{Stella} \&
  {Vietri}}{1998}]{Stella-Vietri98}
{Stella} L.,  {Vietri} M.,  1998, \mn@doi [\apjl] {10.1086/311075}, \href
  {http://adsabs.harvard.edu/abs/1998ApJ...492L..59S} {492, L59}

\bibitem[\protect\citeauthoryear{{Strohmayer}}{{Strohmayer}}{2001a}]{Strohmayer2001b}
{Strohmayer} T.~E.,  2001a, \mn@doi [\apjl] {10.1086/320258}, \href
  {https://ui.adsabs.harvard.edu/abs/2001ApJ...552L..49S} {552, L49}

\bibitem[\protect\citeauthoryear{{Strohmayer}}{{Strohmayer}}{2001b}]{Strohmayer2001}
{Strohmayer} T.~E.,  2001b, \mn@doi [\apjl] {10.1086/321720}, \href
  {https://ui.adsabs.harvard.edu/abs/2001ApJ...554L.169S} {554, L169}

\bibitem[\protect\citeauthoryear{{Sunyaev} \& {Titarchuk}}{{Sunyaev} \&
  {Titarchuk}}{1980}]{Sunyaev1980}
{Sunyaev} R.~A.,  {Titarchuk} L.~G.,  1980, \aap, \href
  {http://adsabs.harvard.edu/abs/1980A%26A....86..121S} {86, 121}

\bibitem[\protect\citeauthoryear{{Tanaka} \& {Lewin}}{{Tanaka} \&
  {Lewin}}{1995}]{Tanaka1995}
{Tanaka} Y.,  {Lewin} W.~H.~G.,  1995, X-ray Binaries, \href
  {http://adsabs.harvard.edu/abs/1995xrbi.nasa..126T} {pp 126--174}

\bibitem[\protect\citeauthoryear{{Varniere}, {Vincent}  \& {Casse}}{{Varniere}
  et~al.}{2020}]{Varniere2020}
{Varniere} P.,  {Vincent} F.~H.,   {Casse} F.,  2020, \mn@doi [\aap]
  {10.1051/0004-6361/202037816}, \href
  {https://ui.adsabs.harvard.edu/abs/2020A&A...638A..33V} {638, A33}

\bibitem[\protect\citeauthoryear{{Vilhu}, {Poutanen}, {Nikula}  \&
  {Nevalainen}}{{Vilhu} et~al.}{2001}]{Vilhu2001}
{Vilhu} O.,  {Poutanen} J.,  {Nikula} P.,   {Nevalainen} J.,  2001, \mn@doi
  [\apjl] {10.1086/320489}, \href
  {https://ui.adsabs.harvard.edu/abs/2001ApJ...553L..51V} {553, L51}

\bibitem[\protect\citeauthoryear{{Vincent}, {Meheut}, {Varniere}  \&
  {Paumard}}{{Vincent} et~al.}{2013}]{Vincent-etal13}
{Vincent} F.~H.,  {Meheut} H.,  {Varniere} P.,   {Paumard} T.,  2013, \mn@doi
  [\aap] {10.1051/0004-6361/201220695}, \href
  {http://adsabs.harvard.edu/abs/2013A%26A...551A..54V} {551, A54}

\bibitem[\protect\citeauthoryear{{Yadav} et~al.,}{{Yadav}
  et~al.}{2016}]{Yadav-etal2016}
{Yadav} J.~S.,  et~al., 2016, \mn@doi [\apj] {10.3847/0004-637X/833/1/27},
  \href {http://adsabs.harvard.edu/abs/2016ApJ...833...27Y} {833, 27}

\bibitem[\protect\citeauthoryear{{Zdziarski}}{{Zdziarski}}{2014}]{Zd2014}
{Zdziarski} A.~A.,  2014, \mn@doi [\mnras] {10.1093/mnras/stu1525}, \href
  {https://ui.adsabs.harvard.edu/abs/2014MNRAS.444.1113Z} {444, 1113}

\bibitem[\protect\citeauthoryear{{Zdziarski}, {Johnson}  \&
  {Magdziarz}}{{Zdziarski} et~al.}{1996}]{Zdziarski_etal1996}
{Zdziarski} A.~A.,  {Johnson} W.~N.,   {Magdziarz} P.,  1996, \mn@doi [\mnras]
  {10.1093/mnras/283.1.193}, \href
  {http://adsabs.harvard.edu/abs/1996MNRAS.283..193Z} {283, 193}

\bibitem[\protect\citeauthoryear{{Zdziarski}, {Grove}, {Poutanen}, {Rao}  \&
  {Vadawale}}{{Zdziarski} et~al.}{2001}]{Zd2001}
{Zdziarski} A.~A.,  {Grove} J.~E.,  {Poutanen} J.,  {Rao} A.~R.,   {Vadawale}
  S.~V.,  2001, \mn@doi [\apjl] {10.1086/320932}, \href
  {https://ui.adsabs.harvard.edu/abs/2001ApJ...554L..45Z} {554, L45}

\bibitem[\protect\citeauthoryear{{Zhang}, {Jahoda}, {Swank}, {Morgan}  \&
  {Giles}}{{Zhang} et~al.}{1995}]{Zhang1995}
{Zhang} W.,  {Jahoda} K.,  {Swank} J.~H.,  {Morgan} E.~H.,   {Giles} A.~B.,
  1995, \mn@doi [\apj] {10.1086/176111}, \href
  {https://ui.adsabs.harvard.edu/abs/1995ApJ...449..930Z} {449, 930}

\bibitem[\protect\citeauthoryear{{van der Klis}}{{van der
  Klis}}{1988}]{VanderKlis1989}
{van der Klis} M.,  1988, in {{\"O}gelman} H.,  {van den Heuvel} E.~P.~J.,
  eds,  NATO Advanced Science Institutes (ASI) Series C Vol. 262, NATO Advanced
  Science Institutes (ASI) Series C. Kluwer Academic Publishers, Dordrecht,
  p.~27

\bibitem[\protect\citeauthoryear{{van der Klis}, {Jansen}, {van Paradijs},
  {Lewin}, {van den Heuvel}, {Trumper}  \& {Szatjno}}{{van der Klis}
  et~al.}{1985}]{VanderKlis1985}
{van der Klis} M.,  {Jansen} F.,  {van Paradijs} J.,  {Lewin} W.~H.~G.,  {van
  den Heuvel} E.~P.~J.,  {Trumper} J.~E.,   {Szatjno} M.,  1985, \mn@doi [\nat]
  {10.1038/316225a0}, \href
  {https://ui.adsabs.harvard.edu/abs/1985Natur.316..225V} {316, 225}

\makeatother
\end{thebibliography}

\label{lastpage}
	
\end{document}